\documentclass[12pt]{article}

\usepackage[utf8]{inputenc}
\usepackage{eurosym,geometry,graphicx,color,setspace,sectsty,comment,footmisc,natbib,pdflscape,array}
\usepackage{caption}
\usepackage[normalem]{ulem}
\usepackage{subcaption}
\usepackage{amssymb,amsfonts,amsmath}
\usepackage{amsthm}
\usepackage{natbib}
\usepackage{booktabs}
\usepackage{threeparttable}
\usepackage{tikzsymbols} 
\usepackage[blocks]{authblk}
\usepackage{makecell}

\usepackage{xcolor} 

\usepackage{multirow}

\newtheorem{assumption}{Assumption}
\newtheorem{lemma}{Lemma}
\newtheorem{theorem}{Theorem}
\newtheorem{corollary}{Corollary}

\theoremstyle{definition} 
\newtheorem{definition}{Definition}

\newtheorem{claim}{Claim}

\newtheorem*{example*}{Example}
\newtheorem{example}{Example}

\usepackage{xr-hyper}
\usepackage{hyperref} 
\hypersetup{colorlinks,citecolor=blue,filecolor=red} 

\title{Identification and Estimation of Dynamic Games \\ with Unknown Information Structure\thanks{We are grateful to Matthew Backus, Gautam Gowrisankaran, Ben Handel, Hidehiko Ichimura, Kei Kawai, Ashley Langer, Sokbae Lee, Qingmin Liu, Juan Pantano, Bernard Salani\'e, Matthijs Wildenbeest, Tiemen Woutersen, Takuro Yamashita, and seminar participants at UArizona, UC Berkeley, Columbia University, and Midwest econometrics group conference. Lorenzo Magnolfi graciously shared his code in the early stage of this project. We merged our two independent working papers previously circulated as ``Estimating Dynamic Games with Unknown Information Structure'' by Koh and ``Estimation of Dynamic Games With Weak Assumptions on Payoff Type Information'' by Hara and Ito. A portion of this paper was completed during Koh's tenure at the U.S. Federal Trade Commission. The views expressed in this article are those of the authors and do not necessarily reflect those of the Federal Trade Commission or any individual Commissioner. All errors are ours.}}
\author{Konan Hara\footnote{Department of Economics, Michigan State University. Email: \texttt{harakon1@msu.edu}} \quad \quad Yuki Ito\footnote{Department of Economics, Indiana University, Bloomington. Email: \texttt{yukiito@iu.edu}} \quad \quad Paul S. Koh\footnote{School of Economics, Yonsei University. Email: \texttt{paulkoh9@gmail.com}}
}

\date{October, 2025}

\begin{document}

\newgeometry{left=1.25in,right=1.25in,top=0.75in,bottom=1.25in}

\maketitle
\begin{abstract} 
\onehalfspacing
   We develop an empirical framework for analyzing dynamic games when the underlying information structure is unknown to the analyst. We introduce \textit{Markov correlated equilibrium}, a dynamic analog of Bayes correlated equilibrium, and show that its predictions coincide with the Markov perfect equilibrium predictions attainable when players observe richer signals than the analyst assumes. We provide tractable methods for informationally robust estimation, inference, and counterfactual analysis. We illustrate the framework with a dynamic entry game between Starbucks and Dunkin' in the US and study the role of informational assumptions.
\end{abstract}  



\restoregeometry

\section{Introduction} \label{sec:introduction} 
This paper develops a computationally tractable econometric framework for estimating empirical dynamic discrete game models with weak assumptions on players' information. Let
\begin{equation}\label{eqn:intro.model}
    f:(\Theta, \mathcal{S}) \rightrightarrows \mathcal{D}
\end{equation}
describe the data-generating process under Markov perfect equilibrium, where $\Theta$ is the set of structural parameters, $\mathcal{S}$ is the set of information structures, and $\mathcal{D}$ is the data; the correspondence in \eqref{eqn:intro.model} reflects the possibility of multiple equilibria. The standard identification argument relies on inverting \eqref{eqn:intro.model} under a specific ``incomplete information'' assumption---namely that players observe their own payoff shocks but not those of their rivals---which simplifies the estimation problem by making it closely parallel to that of single-agent models \citep*{Rust1994, Aguirregabiria2007, Bajari2007, Pakes2007, pesendorfer2008asymptotic, bugni2021iterated, dearing2024efficient}.\footnote{See \citet*{aguirregabiria2021dynamic} for a recent survey of the literature.} 

While this assumption has enabled computationally tractable estimation of dynamic games, researchers often have little evidence to justify it. Unlike observable outcomes such as prices, sales, or entry decisions, information sets are inherently unobservable and may reflect complex mixtures of private knowledge, noisy signals, and shared information transmitted through suppliers, trade associations, or industry channels. Firms may also differ in their ability to gather market intelligence: some obtain precise signals about profitability, while others rely on noisy information. In addition, the timing and granularity of what firms know about their rivals can vary widely. 

A growing body of research shows that assuming the information structure is \emph{known}, while misspecifying players' information, can generate substantial bias in both static (\citealp{grieco2014discrete}; \citealp{pakes2015moment}; \citealp*{Syrgkanis2021};  \citealp{Magnolfi2023}; \citealp{koh2023stable}; \citealp{Gualdani2024}) and dynamic settings (\citealp{Fershtman2012};  \citealp{Aguirregabiria2020}; \citealp*{Asker2020}). The concern is particularly acute in the empirical literature on dynamic games, where most works adopt a specific assumption that limits the role of players' private information in strategic interactions. With the exception of \citet{Pakes1995}, which assumes complete information, we are not aware of empirical approaches that allow for alternative information structures. Achieving informationally robust identification by directly recovering the model primitives from equation \eqref{eqn:intro.model} is, however, infeasible: the set of possible information structures is vast and infinite-dimensional, and the task is further complicated by the inherent challenges of estimating dynamic models of strategic interaction.

In this paper, we develop a tractable empirical framework for analyzing dynamic discrete games while leaving the within-stage information structure \emph{unknown} to the analyst. We assume that the researcher knows only the minimal information available to players but remains agnostic about the possibility that the players may have access to additional signals.\footnote{This assumption restricts the set of possible information structure $\mathcal{S}$ in \eqref{eqn:intro.model}.} The framework typically delivers a set of partially identified parameters that are robust to misspecification of the within-stage information structure. It can be used to construct confidence sets for model parameters that are robust to informational assumptions, thereby assessing the sensitivity of parameter estimates to the researcher's assumptions about players' information.

The main innovation of this paper is the development of \emph{Markov correlated equilibrium}, a dynamic (Markovian) analog of Bayes correlated equilibrium \citep{Bergemann2016}. We show that Markov correlated equilibrium can capture the set of Markov perfect equilibrium predictions (the joint distribution of actions and states) that can arise when the players might observe more signals than assumed by the analyst. We establish the informational robustness property of Markov correlated equilibrium by formulating Markov perfect equilibrium as the Bayes Nash equilibrium of the associated reduced one-shot game via the one-shot deviation principle and invoking Theorem 1 of \citet{Bergemann2016}.

We build on the theory of Markov correlated equilibrium to propose computationally tractable approaches for estimation and inference. Roughly speaking, we prove
\begin{equation}\label{eqn:intro.identified.set}
    \Theta_I^\mathit{MCE} = \bigcup_{\tilde{S} \in \mathcal{S}} \Theta_I^\mathit{MPE} \left(\tilde{S}\right),
\end{equation}
where $\Theta_I^\mathit{MCE}$ is the identified set of parameters after assuming that a Markov correlated equilibrium generated the data, and $\Theta_I^\mathit{MPE}(\tilde{S})$ is the Markov perfect equilibrium identified set when the information structure is assumed to be $\tilde{S}$. Directly inverting \eqref{eqn:intro.model} requires constructing the right-hand side of \eqref{eqn:intro.identified.set}, which is computationally infeasible when $\mathcal{S}$ is large. However, we show that constructing the left-hand side of \eqref{eqn:intro.identified.set} is computationally tractable. In sum, treating the data \emph{as if} they were generated by a Markov correlated equilibrium enables an informationally robust econometric analysis. The identified set \eqref{eqn:intro.identified.set} is usually non-singleton. We propose a computationally attractive strategy for estimation and inference by formulating the estimation problem as mathematical programs with equilibrium constraints (MPEC) \citep{su2012constrained} and applying the inference strategy of \citet{HorowitzLee2022} by nesting the confidence set for the conditional choice probabilities into the MPEC problem.\footnote{The standard estimation strategies for dynamic games, such as the nested fixed-point algorithm of \citet{rust1987optimal} or the conditional choice probability inversion of \citet{hotz1993conditional}, are not applicable.}

We apply our framework to study the dynamic entry and exit decisions of the two largest coffee chains in the US, Starbucks and Dunkin'. As a benchmark, we first estimate the structural parameters under Markov perfect equilibrium, assuming that each player observes their own idiosyncratic payoff shock but not the opponent's, which corresponds to the standard ``incomplete information'' assumption in the literature. We then employ our Markov correlated equilibrium framework to estimate an identified set that captures all possible parameters that could generate the data under Markov perfect equilibrium when the players may additionally observe information about the opponent's payoff shock. We also estimate an identified set that allows for the possibility that the players may not even observe their payoff shock; in this case, $\mathcal{S}$ spans the entire set of information structures, and the identified set is fully robust to any potential misspecification of the information structure. 


Comparison between the estimates from the standard incomplete information Markov perfect equilibrium assumption and our informationally robust Markov correlated equilibrium assumption suggests that the standard incomplete information assumption may be too strong. The estimates from the standard approach suggest that the presence of an opponent has a positive spillover effect on profitability. On the other hand, the Markov correlated equilibrium identified sets do not rule out both positive and negative spillover effects. 

We conduct a counterfactual experiment in which we reduce the firms' entry costs by 10\% and simulate the evolution of the number of active firms, averaged over all markets. The possible outcomes range from 0 (no firms active in any market) to 2 (both firms active in all markets). To illustrate the range of counterfactual predictions under our framework, we consider two extreme scenarios. The low (resp. high) scenario considers the expected number of active firms in equilibrium using the lower (resp. upper) bounds of the 95\% CIs of the competitive effects parameters, while holding all other parameters at their point estimates.

The standard Markov perfect equilibrium approach predicts that reducing entry costs would raise the average number of active firms from the baseline of 0.37 to between 0.55 (low) and 0.58 (high) by the end of the sample period. Allowing for the possibility that firms observe not only their own but also their rival's payoff shock expands the range to 0.35--0.87 under our Markov correlated equilibrium framework. The range becomes even wider, 0.26--1.32, once we incorporate the uncertainty of the remaining structural parameters, and it grows further if we allow for the possibility that firms may not observe (or may only partially observe) their own payoff shocks. 

These results show that our framework can generate policy implications that are both robust to informational assumptions and informative, though the degree of informativeness depends on how flexibly one models the information structure and accounts for parameter uncertainty. At the same time, the robustness provided by Markov correlated equilibrium may admit too many possibilities to be empirically useful, a limitation also noted by \citet{koh2023stable}, \citet{Magnolfi2023}, and \citet{Gualdani2024} in their application of Bayes correlated equilibrium.

\subsubsection*{Related Literature}
Our work advances the literature on the empirical analysis of game-theoretic models in several ways. Our work is most closely related to the econometric frameworks that pursue robustness to misspecification on information structures (\citealp{grieco2014discrete}; \citealp{pakes2015moment}; \citealp*{Syrgkanis2021};   \citealp{koh2023stable}; \citealp{Magnolfi2023}; \citealp{canen2024decomposition}; \citealp{Gualdani2024}; \citealp*{han2024testing}). Building on this literature, we develop a novel framework for estimating a class of dynamic games. More broadly, our work contributes to the literature on tractable econometric methods for partially identified game-theoretic models under weak assumptions on equilibrium selection \citep{Tamer2003, ciliberto2009market, beresteanu2011sharp, galichon2011set}.

We contribute to the dynamic games literature by developing a new solution concept with attractive computational properties. Markov correlated equilibrium extends Bayes correlated equilibrium of \citet{Bergemann2013, Bergemann2016} to dynamic Markovian games in the tradition of \citet{maskin2001markov}. While related notions of correlated equilibrium in dynamic games have been studied \citep{Nowak1992, Duffie1994, solan2001characterization, solan2002correlated, mailath2006repeated, von2008extensive, He2017, forges2020correlated}, our formulation is distinct in extending Bayes correlated equilibrium to dynamic stochastic Markov games with incomplete information. Importantly, our solution concept is tailored to the class of dynamic games that serve as the workhorse for empirical studies of dynamic competition \citep{aguirregabiria2021dynamic}.

Our framework maintains tractability by focusing on robustness to within-stage information assumptions. Although one could allow for across-stage incomplete information and dynamic belief updating (e.g., \citealp{Doval2020}; \citealp{makris2023information}), such extensions are typically infeasible in empirical work. Persistent information asymmetries render estimation intractable \citep{Fershtman2012}, and broadening the solution concept would further enlarge the identified set, reducing its empirical informativeness. Our results already demonstrate that Markov correlated equilibrium admits a wide range of outcomes, which can significantly weaken empirical content, as observed by \citet{koh2023stable}, \citet{Magnolfi2023}, and \citet{Gualdani2024} in static settings. For this reason, we restrict attention to the informational assumptions most relevant to existing empirical frameworks, rather than pursuing broader forms of robustness.


Finally, this paper contributes to the empirical literature on dynamic competition and spatial expansion of retail chains \citep*{Aguirregabiria2007, jia2008happens, holmes2011diffusion, ellickson2013estimating, nishida2015estimating, aguirregabiria2016empirical, arcidiacono2016estimation, yang2020learning, beresteanu2024dynamics}. We examine a novel application: the dynamic expansion of the two largest coffee chains in the US, Starbucks and Dunkin'. The study most closely related to ours is \citet{fang2024measuring}, who analyze entry-deterrence motives of the four largest coffee chains in Toronto, Canada, using a dynamic oligopoly framework.

The rest of the paper is organized as follows. In Section \ref{sec:theory}, we define Markov correlated equilibrium and establish a connection between Markov correlated equilibria and Markov perfect equilibria in a class of dynamic games and discuss its properties. In Section \ref{sec:identification}, we introduce an econometric model and discuss informationally robust identification. Section \ref{sec:estimation} proposes strategies for computing the identified set. Section \ref{sec:inference} discusses our approach to inference. Section \ref{sec:empirical.application} applies our framework to the dynamic entry/exit game by Starbucks and Dunkin' in the US. Section \ref{sec:conclusion} concludes. All proofs are in Appendix \ref{sec:Proofs}.

\section{Theory of Markov Correlated Equilibrium} \label{sec:theory}

We consider a class of dynamic Markovian games that has been used as a standard framework for empirical works. We introduce the concept of Markov correlated equilibrium for informationally robust analysis and discuss its properties.

\subsection{Setup}
Let $t=1,2,...,\infty$ denote discrete time. A \textit{stationary dynamic Markov game of incomplete information} is defined by a pair of basic game $G$ and information structure $S$. A basic game $G = \langle \mathcal{I}, (\mathcal{A}_i, u_i)_{i \in \mathcal{I}}, \mathcal{X}, \mathcal{E}, \psi, f, \delta \rangle$ specifies the payoff-relevant primitives of the model: $i \in \mathcal{I} = \{1, 2, ..., I \}$ indexes the players; $a_{it} \in \mathcal{A}_i$ denotes player $i$'s action at time $t$, where $\mathcal{A}_i$ is a finite set of actions available to player $i$; $a_t = (a_{1t}, ..., a_{It}) \in \mathcal{A} \equiv \times_{i \in \mathcal{I}} \mathcal{A}_i$ denotes an action profile; $x_t \in \mathcal{X}$ denotes a state variable that is publicly observed by the players, and $\mathcal{X}$ is assumed to be finite; $\varepsilon_t \in \mathcal{E}$ denotes a latent state variable that is not directly observed by the players, and $\mathcal{E}$ is assumed to be finite\footnote{The assumption of a finite state space is non-substantive: it simplifies notation and matches the discretization used in estimation. The theory readily extends to a more general state space; see \citet*{Syrgkanis2021} and \citet{Magnolfi2023} for the case of Bayes correlated equilibrium.};  $\psi_{\varepsilon_t \vert x_t}$ represents the players' prior on the probability of $\varepsilon_t$ conditional on $x_t$; $f_{x_{t+1} \vert a_t, x_t, \varepsilon_t}$ specifies the probability of transitioning to state $x_{t+1}$ conditional on $(a_t, x_t, \varepsilon_t)$; $u_i: \mathcal{A} \times \mathcal{X}\times \mathcal{E} \to \mathbb{R}$ is the flow payoff function of player $i$; $\delta \in [0, 1)$ is the players' discount factor. 

An information structure $S = \left( (\mathcal{T}_i)_{i\in \mathcal{I}} , \pi \right)$ specifies the information-relevant primitives; $\tau_{it} \in \mathcal{T}_i$ denotes player $i$'s private signal at period $t$, where $\mathcal{T}_i$ represents a finite set of private signals; $\tau_t = (\tau_{1t}, ..., \tau_{It}) \in \mathcal{T} \equiv \times_{i \in \mathcal{I}} \mathcal{T}_i$ is a signal profile; $\pi : \mathcal{X} \times \mathcal{E} \to \Delta (\mathcal{T})$ is a signal distribution that maps the state of the world to a signal profile. Note that players' signals are allowed to be correlated. The interpretation is that each player $i$ does not directly observe the latent state $\varepsilon_t$ but receives a private signal $\tau_{it}$ whose informativeness about $\varepsilon_t$ depends on $\pi$. 

The game $(G,S)$ is common knowledge to the players. The model proceeds as follows. At the beginning of period $t$, the common knowledge state $x_t \in \mathcal{X}$ is given and publicly observed. Conditional on $x_t$, the latent state $\varepsilon_t \in \mathcal{E}$ is drawn from the probability distribution $\psi_{\varepsilon_t \vert x_t}$. Next, a profile of private signals $\tau_t = (\tau_{1t}, ..., \tau_{It}) \in \mathcal{T}$ is drawn from the signal distribution $\pi_{\tau_t \vert x_t, \varepsilon_t}$. At this point, each player $i$ observes $(x_t, \tau_{it})$ and uses Bayes' rule to infer $\varepsilon_t$. Then, the players simultaneously choose actions $a_{it}$, $i=1,...,I$, and each player $i$ receives period $t$ payoff $u_i(a_t, x_t, \varepsilon_t) \in \mathbb{R}$. Finally, the observable state $x_t$ transitions to $x_{t+1}$ via the probability kernel $f_{x_{t+1} \vert a_t, x_t, \varepsilon_t}$. Given $x_{t+1}$, period $t+1$ begins.

Note that the primitives imply that the transition probability of state variables factors as $\mathrm{Pr}(x_{t+1}, \varepsilon_{t+1} \vert a_t, x_t, \varepsilon_t) = \psi_{\varepsilon_{t+1} \vert x_{t+1}} f_{x_{t+1} \vert a_t, x_t, \varepsilon_t}$. Assuming that the latent variables are generated independently of the previous-period states is standard in the empirical literature.\footnote{When latent variables are correlated over time, empirical analysis quickly becomes intractable, as computation of perfect Bayesian equilibria requires tracking the full history of players' information and updating beliefs at every contingency \citep{Fershtman2012}.} The assumption also plays a vital role in this paper as it simplifies the characterization of correlated equilibria in dynamic settings. 

The players are rational and forward-looking. In period $t$, each player $i$ chooses action plans to maximize the expected discounted sum of intertemporal payoffs
\begin{equation*}
\mathbb{E} \bigg[ \sum_{s=t}^\infty \delta^{s-t} u_i(a_s, x_s, \varepsilon_s) \vert x_t, \tau_{it} \bigg].
\end{equation*}
The conditioning on $(x_t, \tau_{it})$ reflects the assumption that $i$ observes the public state $x_t$ and her signal $\tau_{it}$ before choosing period $t$ action. Since the environment is stationary, we suppress time subscripts unless necessary.

\begin{example}[Two-player dynamic entry game] \label{example:running.example}
As a running example, we consider the two-player dynamic entry/exit model of \citet{pesendorfer2008asymptotic}, which has served as a representative example in the literature (\citealp*{egesdal2015estimating}; \citealp{aguirregabiria2021dynamic}; \citealp{dearing2024efficient}). We also adopt a similar model in our empirical application. There are two players, $i=1,2$. In each period, firms simultaneously decide whether to be active ($a_{it}=1$) or inactive ($a_{it}=0$). The per-period payoff is
\begin{equation}\label{eqn:psd.example.payoff}
    u_i(a_{it},a_{jt},z_{it},\varepsilon_{it}) = 
\begin{cases}
    (1-a_{jt}) \pi^m + a_{jt} \pi^d + (1-z_{it})c + \varepsilon_{it} & \text{if $a_{it} = 1$} \\
    z_{it}\kappa & \text{if $a_{it} = 0$}
\end{cases}
    .
\end{equation}
Here, $\pi^m$, $\pi^d$, $c$, and $\kappa$ represent monopoly profit, duopoly profit, entry cost, and scrap value, respectively. The public state variable is $x_{t}=(z_{1t},z_{2t})$, where $z_{it}$ represents whether firm $i$ is an incumbent ($z_{it}=1$) or not ($z_{it}=0$). The firms' previous actions determine the incumbency states as $z_{it} = a_{i,t-1}$. In empirical applications, $x_{t}$ usually contains additional observable market characteristics. The latent state variable is $\varepsilon_t = (\varepsilon_{1t},\varepsilon_{2t})$, where $\varepsilon_{it} \in \mathcal{E}_i \subseteq \mathbb{R}$ only enters player $i$'s payoff. Each $\varepsilon_{it}$ independently follows the standard normal distribution.   

Standard empirical dynamic discrete game models assume that each player observes $\varepsilon_{it}$ but not $\varepsilon_{jt}$ for $j \neq i$. Formally, $\mathcal{T}_i = \mathcal{E}_i$, and the private signal is $\tau_{it} = \varepsilon_{it}$ with probability one. Alternatively, if the information is complete, players publicly observe $\varepsilon_t$, so $\mathcal{T}_i = \mathcal{E}$, and $\tau_{it} = \varepsilon_t$ with probability one. 
\end{example}

\subsection{Markov Perfect Equilibrium}
\label{sec:markov.perfect.equilibrium}
A Markov strategy of player $i$ is a mapping $\beta_i: \mathcal{X} \times \mathcal{T}_i \to \Delta (\mathcal{A}_i)$ that specifies a probability distribution over actions at each realization of observable state and private signal.\footnote{
While the standard Markov strategy assumption only allows players to condition on payoff-relevant state \citep{maskin2001markov}, our definition allows players' strategy to depend on signals that may be unrelated to the fundamentals (e.g., a public correlation device under complete information), which is a consequence of considering arbitrary information structures. However, our assumptions align with the spirit of \citet{maskin2001markov} because we rule out arbitrary strategies by requiring actions to depend only on signals of a memoryless latent state. Moreover, we are unaware of a treatment for
the construction of the maximally coarse partition of the action space in an incomplete information setup.
} A Markov strategy profile $\beta = (\beta_1, ..., \beta_I)$ is a \textit{Markov perfect equilibrium} if it is a subgame perfect equilibrium.\footnote{This definition is standard in the literature. Beliefs are determined by the signal distribution $\pi$, conditional on publicly observed states. With public transitions and common knowledge of probabilities, beliefs need not enter the equilibrium definition. Subgames are well-defined under Markov strategies.} 

To leverage the results in static games to a dynamic environment, it is useful to characterize the Markov perfect equilibrium conditions using the one-shot deviation principle.\footnote{According to the one-shot deviation principle, a strategy profile is subgame perfect precisely when no player can gain from deviating at a single information set. For infinite-horizon games, this principle holds under the “continuity at infinity” condition, which ensures that far-off events matter little \citep{fudenberg1991game}. Dynamic discrete game models satisfy this condition because payoffs are discounted sums of uniformly bounded expected stage payoffs, even if shocks themselves have unbounded support.} The one-shot deviation principle formulation has two advantages. First, we can leverage results established in static environments as it translates a Markov perfect equilibrium as a Bayes-Nash equilibrium of the associated normal-form game. Second, it facilitates equilibrium computation during econometric analysis.

Suppose a strategy profile $\beta = (\beta_1, ..., \beta_I)$ is given. Define the \textit{ex-ante value function} $V_i^\beta(x) \in \mathbb{R}$ as the expected payoff to player $i$ at state $x$ prior to the realization of $\varepsilon$ when all players follow the prescriptions in $\beta$. For each $i \in \mathcal{I}$, $V_i^\beta \in \mathbb{R}^{\vert \mathcal{X} \vert}$ is the unique solution to
\begin{equation}\label{eqn:mpe.ex.ante.value.function}
V_i^\beta(x) = \sum_{\varepsilon \in \mathcal{E}, \tau \in \mathcal{T}, a \in \mathcal{A}} \psi_{\varepsilon \vert x } \pi_{\tau \vert x, \varepsilon} \beta_{a \vert x, \tau} \bigg \{ u_i(a,x,\varepsilon) + \delta \sum_{x' \in \mathcal{X}} V_i^\beta(x') f_{x' \vert a, x, \varepsilon} \bigg \}, \quad \forall x \in \mathcal{X}
\end{equation}where $\beta_{a \vert x, \tau} \equiv \prod_{i=1}^I \beta_i (a_i \vert x, \tau_i)$ denotes the conditional distribution over action profiles induced by the strategy profile $\beta$. Next, define the \textit{outcome-specific payoff function} given strategy profile $\beta$ as
\begin{equation*}
v_i^\beta(a,x,\varepsilon) \equiv u_i(a,x,\varepsilon) + \delta \sum_{x' \in \mathcal{X}} V_i^\beta(x') f_{x'\vert a, x, \varepsilon}, 
\end{equation*}
which represents the continuation payoff to player $i$ when $(a,x,\varepsilon)$ is realized today, and all players follow the prescriptions in $\beta$ from tomorrow onward.

A strategy profile $\beta$ in game $(G,S) $ induces a reduced normal-form basic game $G^\beta = \langle \mathcal{I}, (\mathcal{A}_i, v_i^\beta )_{i \in \mathcal{I}}, \mathcal{X}, \mathcal{E}, \psi \rangle$. Then $(G^\beta, S)$ describes a ``static" game in which player $i$'s ``static'' payoff function is given by $v_i^\beta : \mathcal{A} \times \mathcal{X} \times \mathcal{E} \to \mathbb{R}$. The following lemma allows us to import the results in static environments into dynamic environments.
\begin{lemma}[One-shot deviation principle formulation of Markov perfect equilibrium] \label{lemma:one.shot.deviation.principle.formulation.of.Markov.perfect.equilibrium}
    A strategy profile $\beta$ is a Markov perfect equilibrium of $(G,S)$ if and only if $\beta$ is a Bayes Nash equilibrium of $(G^\beta, S)$. 
\end{lemma}



\subsection{Markov Correlated Equilibrium}

We introduce a dynamic analog of Bayes correlated equilibrium for dynamic discrete games. A \textit{stationary Markov decision rule} in $(G,S)$ is a mapping 
\begin{equation*}
\sigma: \mathcal{X} \times \mathcal{E} \times \mathcal{T} \to \Delta (\mathcal{A}),
\end{equation*}
which specifies a probability distribution over action profiles at each realization of state and signals. It is instructive to think of $\sigma$ as a recommendation strategy of an omniscient mediator who observes $(x, \varepsilon, \tau)$ and privately recommends action to each player. Suppose the mediator commits to a recommendation strategy $\sigma$ and announces it to the players. After the state and players' signal $(x, \varepsilon, \tau)$ are realized, an action profile $a=(a_1,...,a_I)$ is drawn from the probability distribution $\sigma_{a \vert x, \varepsilon, \tau}$, and each $a_i$ is privately recommended to each player $i$. Each player $i$, having observed $(x, \tau_i, a_i)$, decides whether to obey (play $a_i$) or not (deviate to $a_i' \neq a_i$). If the decision rule $\sigma$ is such that the players are always obedient, we call $\sigma$ a Markov correlated equilibrium of $(G,S)$. 

We formalize the equilibrium condition as follows. Let $V_i^\sigma \in \mathbb{R}^{\vert \mathcal{X} \vert}$ denote the vector of ex-ante value functions induced by $\sigma$, obtained as a unique solution to
\begin{equation}\label{eqn:ex.ante.value.function}
V_i^\sigma(x) = \sum_{\varepsilon \in \mathcal{E}, \tau \in \mathcal{T}, a \in \mathcal{A}} \psi_{\varepsilon \vert x} \pi_{\tau \vert x, \varepsilon} \sigma_{a \vert x, \varepsilon, \tau} \bigg\{ u_i(a,x,\varepsilon) + \delta \sum_{x' \in \mathcal{X}} V_i^\sigma (x') f_{x' \vert a, x, \varepsilon} \bigg \}, \quad \forall x \in \mathcal{X}.
\end{equation}
Define the outcome-specific value function associated with $\sigma$ as
\begin{equation*}
v_i^\sigma (a,x,\varepsilon) \equiv u_i(a,x,\varepsilon) + \delta \sum_{x' \in \mathcal{X}} V_i^\sigma(x') f_{x' \vert a, x, \varepsilon},
\end{equation*}
which represents the payoff to player $i$ if $(a,x,\varepsilon)$ is realized today and $\sigma$ determines the players' actions in the future. Let $\mathbb{E}^\sigma [ v_i^\sigma (a_i', a_{-i}, x, \varepsilon) \vert x, \tau_i, a_i]$ be the expected payoff to player $i$ from choosing $a_i'$ when $i$ observes $(x,\tau_i)$ and receives recommendation $a_i$. The following definition states that $\sigma$ is a Markov correlated equilibrium of $(G,S)$ if the players do not deviate from the mediator's recommendations.
\begin{definition}
    A decision rule $\sigma$ is a \textit{Markov correlated equilibrium} of $(G,S)$ if for each $i \in \mathcal{I}$, $x\in \mathcal{X}$, $\tau_i \in \mathcal{T}_i$, and $a_i \in \mathcal{A}_i$, we have
    \begin{equation} \label{eqn:MCEobedience.original}
    \mathbb{E}^\sigma [ v_i^\sigma(a_i,a_{-i},x,\varepsilon) \vert x, \tau_i, a_i] \geq \mathbb{E}^\sigma [ v_i^\sigma(a_i',a_{-i},x,\varepsilon) \vert x, \tau_i, a_i ]
    \end{equation}
    for each $a_i' \neq a_i$ whenever $\mathrm{Pr}^\sigma (x,\tau_i,a_i) >0$.
\end{definition}

Since 
\begin{align*}
\mathbb{E}^\sigma[v_i^\sigma(\tilde{a}_i, a_{-i}, x, \varepsilon) \vert x, \tau_i, a_i ] &= \sum_{\varepsilon, a_{-i}} v_i^\sigma(\tilde{a}_i, a_{-i}, x, \varepsilon) \mathrm{Pr}^\sigma (\varepsilon, a_{-i} \vert x, \tau_i, a_i) \\
& = \sum_{\varepsilon, a_{-i}} v_i^\sigma (\tilde{a}_i, a_{-i},x,\varepsilon) \Bigg( \frac{\sum_{\tau_{-i}} \psi_{\varepsilon \vert x} \pi_{\tau \vert x, \varepsilon} \sigma_{(a_i, a_{-i}) \vert x, \varepsilon, \tau} }{\sum_{\bar{\varepsilon}, \tau_{-i}, \bar{a}_{-i}} \psi_{\bar{\varepsilon} \vert x} \pi_{\tau \vert x, \bar{\varepsilon}} \sigma_{(a_i, \bar{a}_{-i}) \vert x, \bar{\varepsilon}, \tau}} \Bigg),
\end{align*}
canceling out the denominator---which is constant across all possible realizations of $(\varepsilon, \tau_{-i}, a_{-i})$---lets us rewrite the obedience condition \eqref{eqn:MCEobedience.original} as, for each $i$,
\begin{equation}\label{eqn:MCEobedience00}
\sum_{\varepsilon, \tau_{-i}, a_{-i}} \psi_{\varepsilon \vert x} \pi_{\tau \vert x, \varepsilon} \sigma_{a \vert x, \varepsilon, \tau} v_i^\sigma(a,x,\varepsilon) \geq \sum_{\varepsilon, \tau_{-i}, a_{-i}} \psi_{\varepsilon\vert x} \pi_{\tau \vert x, \varepsilon} \sigma_{a \vert x, \varepsilon, \tau} v_i^\sigma (a_i', a_{-i}, x, \varepsilon), \quad \forall x, \tau_i , a_i, a_i'.
\end{equation}
Equation \eqref{eqn:MCEobedience00} serves as the operational definition of Markov correlated equilibrium in our econometric analysis.

A Markov correlated equilibrium always exists for any signal distribution $\pi$. \citet{Nowak1992} establishes the existence of a stationary correlated equilibrium in a discounted stochastic game with complete information under standard regularity conditions. We will show in the following sections that the set of Markov correlated equilibria under arbitrary information structure always includes a complete information stationary correlated equilibrium.\footnote{Existence results for Markov perfect equilibrium under incomplete information do not necessarily imply the existence of a Markov correlated equilibrium for certain information structures. For this reason, we rely on the result of \citet{Nowak1992}.} Thus, the set of Markov correlated equilibria is always non-empty.

\subsection{Relationship to Bayes Correlated Equilibrium}

As in Section \ref{sec:markov.perfect.equilibrium}, a decision rule $\sigma$ in game $(G,S)$ induces a reduced normal-form basic game $G^\sigma = \langle \mathcal{I}, (\mathcal{A}_i, v_i^\sigma)_{i\in \mathcal{I}}, \mathcal{X}, \mathcal{E}, \psi \rangle$. Then \eqref{eqn:MCEobedience00} shows that if $\sigma$ is a Markov correlated equilibrium of $(G,S)$, it is a Bayes correlated equilibrium of $(G^\sigma,S)$. 
\begin{lemma}[One-shot deviation principle characterization of Markov correlated equilibrium] \label{Lemma:MCEandBCE}
    A decision rule $\sigma$ is a Markov correlated equilibrium of $(G,S)$ if and only if it is a Bayes correlated equilibrium of $(G^\sigma,S)$. 
\end{lemma}

When players fully discount the future ($\delta = 0$), a Markov correlated equilibrium collapses to a Bayes correlated equilibrium of the stage game: for each $i$,
\begin{equation*}
\sum_{\varepsilon, \tau_{-i}, a_{-i}} \psi_{\varepsilon \vert x} \pi_{\tau \vert x, \varepsilon} \sigma_{a \vert x, \varepsilon, \tau} u_i(a,x,\varepsilon) \geq \sum_{\varepsilon, \tau_{-i}, a_{-i}} \psi_{\varepsilon\vert x} \pi_{\tau \vert x, \varepsilon} \sigma_{a \vert x, \varepsilon, \tau} u_i (a_i', a_{-i}, x, \varepsilon), \quad \forall x, \tau_i , a_i, a_i'.
\end{equation*}
Thus, Markov correlated equilibrium is a proper dynamic analog of Bayes correlated equilibrium.

Unfortunately, contrary to its static analog, the Markov correlated equilibrium conditions are not linear with respect to the decision rule, which creates two challenges that are absent in the static case. First, Markov correlated equilibrium is more difficult to compute. While computing a Bayes correlated equilibrium amounts to solving a linear program, computing a Markov correlated equilibrium generally requires solving a non-convex program. Second, the set of Bayes correlated equilibria is convex, but the set of Markov correlated equilibria is generally non-convex. The convexity of the set of Bayes correlated equilibria implies that a mixture of multiple equilibria is observationally equivalent to a single equilibrium; \citet*{Syrgkanis2021} and \citet{Magnolfi2023} use this property to stay agnostic about the true equilibrium selection rule in the data generating process. However, since Markov correlated equilibrium does not share this property, we assume that a single equilibrium is selected.

\subsection{Informational Robustness}
We establish the informational robustness property of Markov correlated equilibrium. Let an equilibrium prediction refer to the joint distribution on actions, states, and signals that are implied by either an equilibrium strategy profile or decision rule. We prove the following claim.
\begin{claim}[Markov correlated equilibrium is informationally robust]\label{claim:informational.robustness.of.mce}
    The set of Markov correlated equilibrium predictions is equal to the set of Markov perfect equilibrium predictions that can arise when players can observe more signals that are unknown to the analyst.
\end{claim}
In the empirical application, we consider a scenario where the analyst knows that each firm observes its firm-specific payoff shock but cannot rule out the possibility that they receive some signals about opponents' shocks. Claim \ref{claim:informational.robustness.of.mce} says that the analyst can capture the set of all predictions that can arise in such a scenario by resorting to Markov correlated equilibrium. Thus, as in Bayes correlated equilibrium, Markov correlated equilibrium attains a  form of informational robustness that allows the analyst to be agnostic about the players observing additional signals after specifying players' minimal information. We formalize Claim \ref{claim:informational.robustness.of.mce} by following \citet{Bergemann2016}'s framework and connecting Markov correlated equilibrium to Bayes correlated equilibrium via the one-shot deviation principle. 

First, we introduce a partial order on the set of information structures using the notion of \emph{expansion} as in \citet{Bergemann2016}. Let $\omega \in \Omega \equiv \mathcal{X}\times \mathcal{E}$ for the sake of exposition.
\begin{definition}[Expansion]
    Let $S = (\mathcal{T},\pi)$ be an information structure. Information structure $S^* = (\mathcal{T}^*, \pi^*)$ is an \textit{expansion} of $S$, or $S^* \succsim_E S$, if there exists $(\tilde{\mathcal{T}_i})_{i\in \mathcal{I}}$ and $\lambda:\Omega \times \mathcal{T} \to \Delta(\tilde{\mathcal{T}})$ such that $\mathcal{T}_i^* = \mathcal{T}_i \times \tilde{\mathcal{T}}_i$ for all $i\in \mathcal{I}$ and $\pi^*_{\tau, \tilde{\tau} \vert \omega} = \pi_{\tau \vert \omega} \lambda_{\tilde{\tau} \vert \omega, \tau}$. 
\end{definition} 
Intuitively, if $S^* \succsim_E S$, the players receive \textit{extra} signals in $S^*$ than they do in $S$. Specifically, complete information (players observe the latent state $\varepsilon$ almost surely) is an expansion of an arbitrary information structure.

Next, we claim that if a strategy profile induces the same joint distribution over states and actions as a decision rule, the two induce the same reduced normal-form games. Let $S^*$ be an expansion of $S$. A strategy profile $\beta$ in $(G,S^*)$ \textit{induces} a decision rule $\sigma$ for $(G,S)$ if
\begin{equation*}
\sigma_{a\vert x, \varepsilon, \tau} = \sum_{\tilde{\tau} \in \tilde{\mathcal{T}}} \lambda_{\tilde{\tau} \vert x, \varepsilon, \tau} \prod_{i=1}^I \beta_i (a_i \vert x, \tau_i, \tilde{\tau}_i), \quad \forall a, x, \varepsilon, \tau.
\end{equation*}
The strategy profile $\beta$ in a game with more signals induces the same joint distribution over actions, states, and signals as the decision rule $\sigma$ in a game with fewer signals.

\begin{lemma}[Identical reduced normal-form basic games] \label{lemma:identical.reduced.form.games}
    If $\beta$ induces $\sigma$, then the associated reduced normal-form basic games, $G^\beta$ and $G^\sigma$, are identical.
\end{lemma}

Lemma \ref{lemma:identical.reduced.form.games} is intuitive: if $\beta$ induces $\sigma$, then $\beta$ and $\sigma$ induce identical distributions over action profiles at each state, so the associated ex-ante value functions must be identical to each other, i.e., $V_i^\beta = V_i^\sigma$. Then, the outcome-specific payoff functions are also identical to each other, i.e., $v_i^\beta = v_i^\sigma$, making all primitives of the basic games identical to each other.

Finally, a solution concept generates a \textit{prediction} defined as a probability distribution over actions at each state and signal. Let $\mathcal{P}_{a \vert x,\varepsilon,\tau}^{\mathit{MPE}}\left(G,S\right)$ represent the set of predictions that can be induced by a Markov perfect equilibrium in $\left(G,S\right)$.\footnote{We slightly abuse the notation and use $\mathcal{P}_{a\vert x, \varepsilon, \tau}^{\mathit{MPE}} (G,S)$ to denote a set of conditional distributions over $\mathcal{A}$ defined at each $(x,\varepsilon,\tau)$.} Let $\mathcal{P}_{a\vert x,\varepsilon,\tau}^{\mathit{MCE}}\left(G,S\right)$ be defined similarly for a Markov correlated equilibrium. 

We are ready to state our main theorem. Suppose the analyst knows that the players observe signals specified by $S$ but does not know whether the players have access to extra signals. What is the set of feasible Markov perfect equilibrium predictions? The following theorem says that the set of Markov correlated equilibria of $(G,S)$ captures all Markov perfect equilibrium predictions when the players can observe more signals than in $S$. 

\begin{theorem}[Markov correlated equilibrium is informationally robust] \label{thm:informational_robustness}
    For any basic game $G$ and information structure $S$, 
    \begin{equation*}
    \mathcal{P}^{\mathit{MCE}}_{a\vert x, \varepsilon, \tau}(G,S) = \bigcup_{S^* \succsim_E S} \mathcal{P}^{\mathit{MPE}}_{a \vert x, \varepsilon, \tau} (G,S^*).
    \end{equation*}
\end{theorem}

Theorem \ref{thm:informational_robustness} is attractive because $\mathcal{P}_{a\vert x,\varepsilon,\tau}^{\mathit{MCE}}\left(G,S\right)$ is far easier to characterize and compute than $\bigcup_{S^{*}\succsim_{E}S}\mathcal{P}_{a\vert x,\varepsilon,\tau}^{\mathit{MPE}}$; the former does not require searching over the set of information structures.\footnote{\citet{makris2023information} obtain a related result in the context of multistage games. However, since their focus is on finite-stage games and they allow for varying information structures \textit{across} stages, our result is not a direct corollary of theirs. Moreover, applying their framework in empirical settings would generally be infeasible, as it requires keeping track of the entire history.} Note that when $\delta = 0$, Theorem \ref{thm:informational_robustness} reduces to
    \begin{equation*}
        \mathcal{P}^{\mathit{BCE}}_{a\vert x, \varepsilon, \tau}(G,S) = \bigcup_{S^* \succsim_E S} \mathcal{P}^{\mathit{BNE}}_{a \vert x, \varepsilon, \tau} (G,S^*),
    \end{equation*}
which is analogous to \cite{Bergemann2016} Theorem 1.

In empirical works, it is common to assume that the analyst observes the \emph{conditional choice probabilities} that represent the probability of each action profile at each state that is observable to the econometrician. Thus, it is useful to characterize the implications of Theorem \ref{thm:informational_robustness} in terms of conditional choice probabilities. Let $\mathcal{P}_{a\vert x}^{\mathit{MPE}}\left(G,S\right)$ denote the set of feasible conditional choice probabilities that can arise under a Markov perfect equilibrium of $\left(G,S\right)$. Let $\mathcal{P}_{a\vert x}^{\mathit{MCE}}\left(G,S\right)$ be defined similarly. 
\begin{corollary} \label{cor:informational.robustness.ccp}
    For any basic game $G$ and information structure $S$, 
    \begin{equation*}
    \mathcal{P}_{a\vert x}^{\mathit{MCE}}\left(G,S\right)=\bigcup_{S^{*}\succsim_{E} S}\mathcal{P}_{a\vert x}^{\mathit{MPE}}\left(G,S^{*}\right).
    \end{equation*}
\end{corollary}

In Section \ref{sec:identification}, we discuss how results from Theorem \ref{thm:informational_robustness} and Corollary \ref{cor:informational.robustness.ccp} can be applied to derive results on the identification of the model parameters.

As in the case of Bayes correlated equilibrium, if players observe more signals, the set of Markov correlated equilibria becomes smaller. In other words, more information implies a smaller set of Markov correlated equilibria.
\begin{theorem} \label{thm:more.info.smaller.mce}
    Let $G$ be an arbitrary basic game. If $S^1 \succsim_E S^2$, then $\mathcal{P}_{a \vert x, \varepsilon, \tau }^{\mathit{MCE} } (G,S^1) \subseteq \mathcal{P}_{a \vert x, \varepsilon, \tau}^{\mathit{MCE}} (G,S^2)$. 
\end{theorem}

\subsection{Numerical Example}

We continue Example \ref{example:running.example} to illustrate the Markov correlated equilibrium identified set. The true parameters in \eqref{eqn:psd.example.payoff} are set to $(\pi^m,\pi^d,c,\kappa)=(1.2,-1.2,-0.2,0.1)$, and we use the discount factor $\delta = 0.9$. We generate data by finding a Markov perfect equilibrium under the standard assumption that $\varepsilon_i$ is private information to player $i$. \citet{pesendorfer2008asymptotic} finds five equilibria, two of which are symmetric to other equilibria. We report the conditional choice probabilities of equilibria (i), (ii), and (iii) of \citet{pesendorfer2008asymptotic} in Table \ref{tab:equilibria.in.psd.game}.\footnote{We use \citet{dearing2024efficient}'s MATLAB code to generate the equilibrium conditional choice probabilities.} 

\begin{table}[htb!]

\caption{\label{tab:equilibria.in.psd.game}Probability of entry in three Markov perfect equilibria}
\centering
\begin{threeparttable}
\begin{tabular}[t]{lcccccc}
\toprule
\multicolumn{1}{c}{ } & \multicolumn{2}{c}{(i)} & \multicolumn{2}{c}{(ii)} & \multicolumn{2}{c}{(iii)} \\
\cmidrule(l{3pt}r{3pt}){2-3} \cmidrule(l{3pt}r{3pt}){4-5} \cmidrule(l{3pt}r{3pt}){6-7}
 & P1 & P2 & P1 & P2 & P1 & P2\\
\midrule
Out/Out & 0.733 & 0.276 & 0.615 & 0.528 & 0.576 & 0.576\\
Out/In & 0.613 & 0.420 & 0.312 & 0.840 & 0.305 & 0.842\\
In/Out & 0.800 & 0.223 & 0.831 & 0.303 & 0.842 & 0.305\\
In/In & 0.752 & 0.294 & 0.606 & 0.578 & 0.595 & 0.595\\
\bottomrule
\end{tabular}
\begin{tablenotes}[para]
\footnotesize
\item \textit{Note:} Each value represents the entry probability of each player at each state. P1 and P2 represent player 1 and player 2.
\end{tablenotes}
\end{threeparttable}
\end{table}

To characterize the bounds on the game outcomes that can arise in a Markov correlated equilibrium, we compute the Markov correlated equilibria that maximize and minimize the expected number of active firms.\footnote{The procedure is similar to the counterfactual equilibrium calculations in the empirical application detailed in Online Appendix \ref{sec:computation.equilibrium}.} Using the associated conditional choice probabilities, we simulate the number of active firms over $T=6$ periods, assuming that the initial state has no incumbent, i.e., $x_1 = (0,0)$.

\begin{figure}[htb!]
    \centering
    \includegraphics[width=0.6\linewidth]{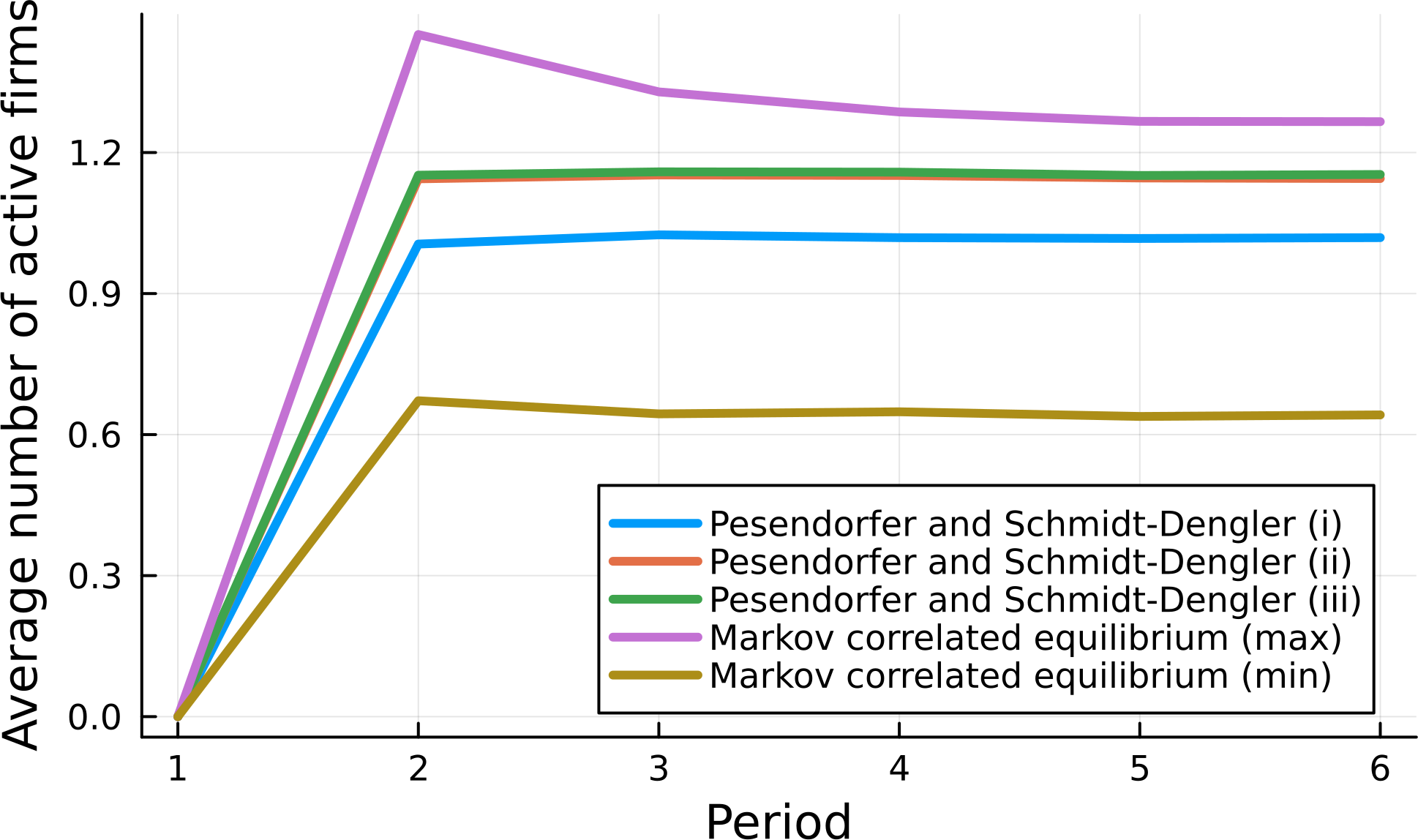}
    \caption{Average number of active firms over time}
    \label{fig:average.number.of.active.firms.over.time}
\end{figure}

Figure \ref{fig:average.number.of.active.firms.over.time} plots the simulated average number of active firms for $T=6$ periods. Note that the outcomes of \citet{pesendorfer2008asymptotic}'s three Markov perfect equilibria are bounded by those of the Markov correlated equilibria. The Markov correlated equilibrium bounds characterize the sharp bounds on the Markov perfect equilibrium outcomes that can arise when players can receive information about the opponent's payoff shock.

\section{Informationally Robust Identification} \label{sec:identification}
In this section, we specialize the model for econometric analysis and discuss informationally robust identification in dynamic discrete games using Markov correlated equilibrium.

\subsection{Setup}
To facilitate econometric analyses, we impose the following assumptions on the basic game primitives. As commonly assumed in empirical works, we let $\varepsilon \equiv (\varepsilon_1, \varepsilon_2, ..., \varepsilon_I)$ where $\varepsilon_i \in \mathcal{E}_i$ only enters player $i$'s payoff and $f_{x'\vert a,x,\varepsilon}=f_{x'\vert a,x}$, i.e., the transition probability of public state $x$ is independent of latent state $\varepsilon$. We assume that the payoff functions and the prior distribution are parameterized by a finite-dimensional vector $\theta$ so that $u_{i}=u_{i}^{\theta}$ and $\psi=\psi^{\theta}$. 

The econometrician observes data $$\left\{ a_{m,t},x_{m,t}:\ m=1,2,...,M,\ t=1,2,...,T\right\},$$ which have players' actions $a_{m,t}$ and common knowledge state variables $x_{m,t}$ across $M$ independent markets over $T$ periods. We assume that $M$ is large and that the conditional choice probabilities $\phi:\mathcal{X}\to\Delta\left(\mathcal{A}\right)$ and the transitional probability function $f:\mathcal{A}\times\mathcal{X}\to\Delta\left(\mathcal{X}\right)$ can be non-parametrically identified from the data. Finally, we assume that the common discount factor $\delta \in [0,1)$ is known to the econometrician. In sum, we treat the basic game $G^\theta = \langle \mathcal{I}, (\mathcal{A}_i, u_i^\theta)_{i\in \mathcal{I}}, \mathcal{X}, \mathcal{E}, \psi^\theta, f, \delta \rangle$ as known to the econometrician up to $\theta$.

We summarize the baseline assumptions for econometric analysis as follows.

\begin{assumption}[Baseline assumptions for econometric analysis] \label{assu:baseline} \phantom{.}

\begin{enumerate}
    \item The set of covariates $\mathcal{X}$ and the set of latent states $\mathcal{E}$ are finite.
    \item The prior distribution $\psi^\theta$ and the payoff functions $u_i^\theta$ are known up to a finite-dimensional parameter $\theta$.
    \item The latent state is a vector of player-specific payoff shocks, i.e., $\varepsilon=\left(\varepsilon_{1},...,\varepsilon_{I}\right)$ and $u_{i}^{\theta}\left(a,x,\varepsilon\right)=u_{i}^{\theta}\left(a,x,\varepsilon_{i}\right)$, and does not affect the transition probability of public states, i.e., $f_{x'\vert a,x,\varepsilon}=f_{x'\vert a,x}$.
    \item The conditional choice probabilities $\phi:\mathcal{X}\to\Delta\left(\mathcal{A}\right)$ and the transition probability function $f:\mathcal{A}\times\mathcal{X}\to\Delta\left(\mathcal{A}\right)$ are identified from the data. 
\end{enumerate}
\end{assumption}

The only ``unconventional" assumption is that $\mathcal{E}$ is finite. Although unnecessary for the identification arguments, the discrete support assumption is necessary for feasible estimation and makes the connection to the previous section straightforward. Other econometric papers that use Bayes correlated equilibrium also discretize the state space for feasible estimation (\citealp*{Syrgkanis2021}; \citealp{Magnolfi2023}; \citealp{Gualdani2024}). The player-specific payoff shock assumption, which is a conventional assumption, makes the role of the latent state clear, but it is also not required for our identification arguments.

Given a game $(G^\theta,S)$, a solution concept determines a set of feasible predictions, which in turn induces a set of feasible conditional choice probabilities. A standard econometric approach is to assume that the information structure $S$ is known and identify the model parameters by inverting the mapping from the model parameters to the observed conditional choice probability vector $\phi$.

\begin{definition}[Identified set]
    Given Assumption \ref{assu:baseline}, a solution concept $\mathit{SC}$, and an information structure $S$, the identified set of parameters is defined as:
    \begin{equation*}
        \Theta_{I}^{\mathit{SC}}\left( S \right)\equiv\left\{ \theta\in\Theta:\ \phi\in\mathcal{P}_{a\vert x}^{\mathit{SC} } (G^{\theta},S ) \right\} .
    \end{equation*}
\end{definition}

In words, a candidate parameter $\theta$ enters the identified set if the observed conditional choice probabilities $\phi$ can be rationalized by some equilibrium (defined by the solution concept) of the model.

\subsection{Informationally Robust Identified Set}
We consider a scenario where a Markov perfect equilibrium generates the data, but the true information structure is unknown to the analyst; misspecifying the information structure leads to biased parameter estimates. We assume that the analyst only knows that the players minimally observe signals in $S$ but may observe more. For instance, as assumed in \citet{Magnolfi2023}, the analyst might know that each player $i$ observes at least $\varepsilon_{i}$ but does not know whether they can receive extra information about opponents' payoff shocks. 

\begin{assumption}[Identification under Markov perfect equilibrium] \label{assu:MPE DGP}
    The data are generated by a Markov perfect equilibrium of $(G^{\theta_0},S^0)$, where $S^{0}$ is an expansion of $S$. 
\end{assumption}

It is well-understood that working directly with Assumption \ref{assu:MPE DGP} is computationally infeasible; when the analyst knows $S$ but does not know the true information structure $S^0$, the analyst has to consider the set of all information structures that are expansions of $S$, but the set is too large. We can make the identification problem computationally feasible by replacing Assumption \ref{assu:MPE DGP} with the following.

\begin{assumption}[Identification under Markov correlated equilibrium] \label{assu:MCE DGP}
    The data are generated by a Markov correlated equilibrium of $\left(G^{\theta_{0}},S\right)$. 
\end{assumption}

\begin{theorem}[Equivalence of identified sets]  \label{thm:equivalence of identified sets}
    The identified set under Assumptions \ref{assu:baseline} and \ref{assu:MPE DGP} is equal to the identified set under Assumptions \ref{assu:baseline} and \ref{assu:MCE DGP}.
\end{theorem}

Theorem \ref{thm:equivalence of identified sets} says that an econometrician who does not know the true information structure underlying the data-generating process can proceed by treating the data \emph{as if} they were generated by a Markov correlated equilibrium. \citet*{Syrgkanis2021}, \citet{Magnolfi2023}, \citet{koh2023stable}, and \citet{Gualdani2024} use similar results in static settings. Theorem \ref{thm:equivalence of identified sets} extends the results to a class of dynamic games. Note that a Markov correlated equilibrium identified set always includes the parameters that can be obtained under the complete information Markov perfect equilibrium assumption with mixed strategies \citep{Pakes1995, Doraszelski2010}.

\subsection{Properties of Informationally Robust Identified Set} \label{sec:properties.informationally.robust.identified.set}

\paragraph{Identified Set Shrinks with Stronger Informational Assumptions}
The Markov correlated equilibrium identified set depends on the baseline information structure set by the analyst, which specifies what players minimally observe. Theorem \ref{thm:more.info.smaller.mce} has shown that a stronger assumption on players' information leads to a tighter set of predictions. The following translates this intuition to identified sets.
\begin{theorem}\label{thm:tighter_set}
    If $S^1 \succsim_E S^2$, then $\Theta_I^{MCE}(S^1) \subseteq \Theta_I^{MCE}(S^2)$. 
\end{theorem}
Theorem \ref{thm:tighter_set} implies a trade-off between informational robustness and the tightness of the identified set. Using weaker assumptions on information is more robust but increases the size of the identified set. The econometrician obtains the tightest identified set when players' baseline information is complete; the econometrician obtains the loosest identified set when players minimally observe nothing about $\varepsilon$.

\paragraph{Markov Correlated Equilibrium is Not Robust to Multiple Equilibria Issue}
In the static case, assuming that a single Bayes correlated equilibrium generates the data is innocuous because any mixture of Bayes correlated equilibria is a Bayes correlated equilibrium. However, this property does not carry over to the dynamic case because a mixture of Markov correlated equilibria need not be a Markov correlated equilibrium. This occurs because the Bayes correlated equilibrium obedience condition is linear in the decision rule while the Markov correlated equilibrium obedience condition is nonlinear (non-convex) in the decision rule. Thus, although we remain agnostic about the equilibrium selection rule, we assume that the data are generated by a single equilibrium. Identification with data generated by multiple equilibria is possible (e.g., by using \citet*{beresteanu2011sharp}), but far more computationally challenging.\footnote{Our framework already incorporates public randomization that is independent across stages. However, because of the relationship between stage-game payoffs and value functions induced by the Markov structure, public randomization does not enable us to convexify the objective set of distributions.}

\paragraph{Excluded Variables with Large Support Ensure Point-Identification} While the tightness of the identified set is largely an empirical question, the identified set can shrink substantially when excluded variables have large support. In a scenario where players know their own payoff shock, \citet{Magnolfi2023} prove that players' payoff parameters are point-identified under Bayes correlated equilibrium when there are player-specific excluded variables with large support.\footnote{Parameter governing the correlation of players' payoff types may not be point-identified.} Consider, for example, a two-player entry game. Driving a time-persistent variable that enters only player $j$'s payoff to either positive or negative infinity ensures that the player always stays in or stays out, rendering the other player's decision problem a single-agent binary choice problem. The econometrician can use the single-agent data to identify the payoff parameters, including the competitive effects parameters \citep{Tamer2003}. The same intuition holds in the dynamic case: in markets where one player always stays in or out, the other player solves a single-agent dynamic discrete choice problem while taking the other player's action as fixed. The players' payoff parameters can be identified and estimated using the single-agent dynamic discrete choice data if our interest lies in the case where players know their own payoff shock.\footnote{If we are interested in the case where players may not have complete knowledge about their own shock, we will need to extend Theorem 1 of \citet{Bergemann2016} to a single-agent dynamic discrete choice setup.}

\subsection{Numerical Example}
Following \citet{pesendorfer2008asymptotic}, we assume that the econometrician knows the values of the scrap value $\kappa=0.1$ and the discount factor $\delta=0.9$. However, instead of assuming that the researcher knows the information structure exactly, we assume that the researcher only knows that $i$ minimally observes $\varepsilon_i$ and leaves open the possibility that $i$ receives extra signals about $\varepsilon_{-i}$. 

Table \ref{tab:mce.identified.sets.numerical.example} reports the projection intervals of the Markov correlated equilibrium identified sets obtained under each equilibrium. In all cases, the Markov correlated equilibrium identified set contains the true parameter vector, as expected. However, the size of the identified set can vary depending on which equilibrium generated data. We find that the Markov correlated equilibrium identified set obtained from equilibrium (i)---which is the only nested pseudo-likelihood-stable equilibrium and generates highly asymmetric choice probabilities across players---is the tightest.\footnote{The nested pseudo-likelihood-stability has been the subject of investigation in the dynamic games estimation literature (see \citet{dearing2024efficient} for a review). Equilibrium (i) is nested pseudo-likelihood-stable while equilibria (ii) and (iii) are not. Our Markov correlated equilibrium estimator is not sensitive to the nested pseudo-likelihood-stability of the underlying equilibrium.} The identified sets under equilibria (ii) and (iii) are indistinguishable because the equilibrium conditional choice probabilities are similar to each other.

\begin{table}[htb!]

\caption{\label{tab:mce.identified.sets.numerical.example}Markov correlated equilibrium identified sets}
\centering
\begin{threeparttable}
\begin{tabular}[t]{lcccc}
\toprule
 & True & (i) & (ii) & (iii)\\
\midrule
$\pi^m$ & 1.2 & $[0.93, 1.83]$ & $[0.88, 2.52]$ & $[0.88, 2.52]$\\
$\pi^d$ & -1.2 & $[-1.87, -1.01]$ & $[-3.16, -0.65]$ & $[-3.16, -0.65]$\\
$c$ & -0.2 & $[-0.51, 0.37]$ & $[-0.67, 1.02]$ & $[-0.67, 1.02]$\\
\bottomrule
\end{tabular}
\begin{tablenotes}[para]
\footnotesize
\item \textit{Note:} The table reports the projections of the Markov correlated equilibrium identified set assuming that the data were generated from \citet{pesendorfer2008asymptotic}'s Markov perfect equilibrium (i), (ii), or (iii).
\end{tablenotes}
\end{threeparttable}
\end{table}

\begin{figure}[htb!]
     \centering
     \begin{subfigure}{0.32\textwidth}
         \centering
         \includegraphics[width=\textwidth]{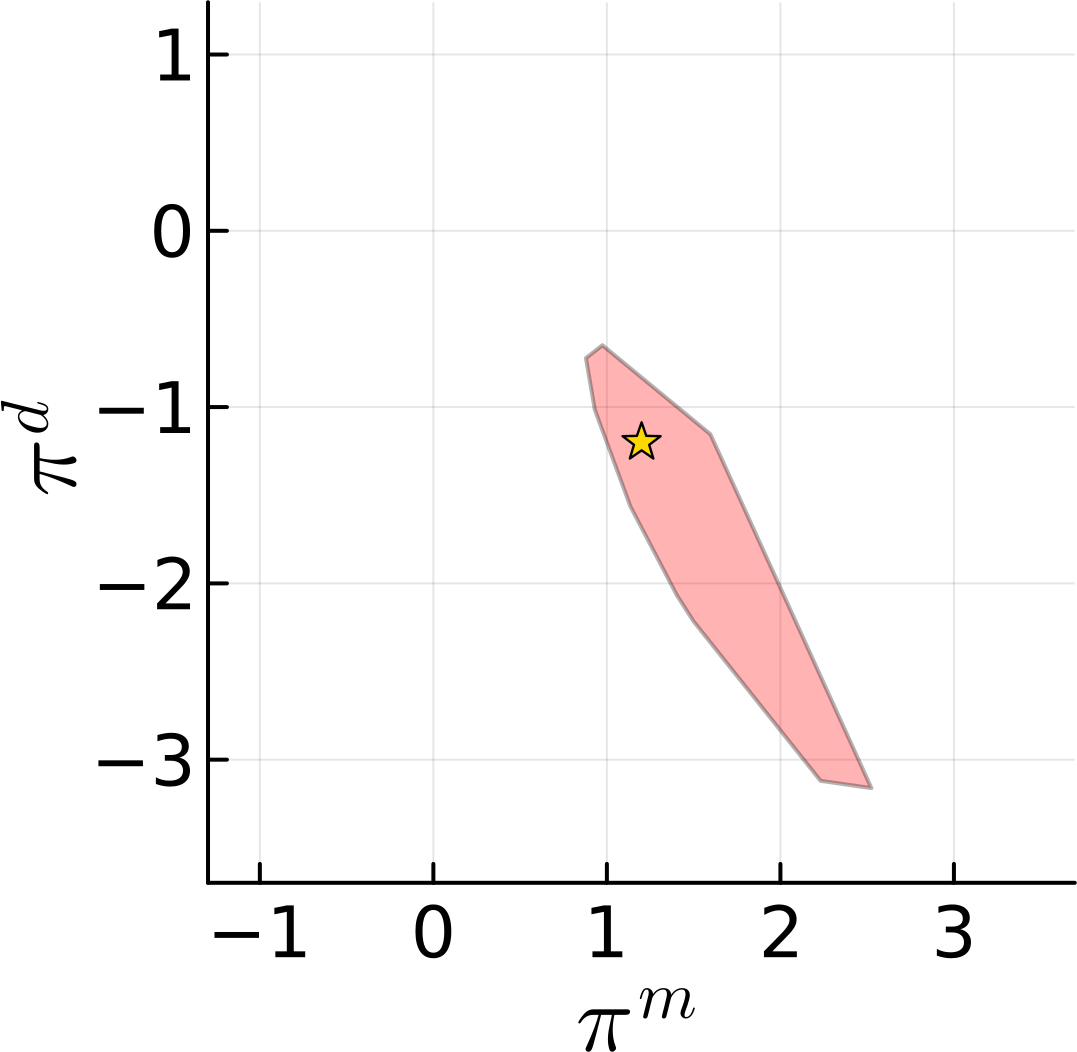}
     \end{subfigure}
     \hfill
     \begin{subfigure}{0.32\textwidth}
         \centering
         \includegraphics[width=\textwidth]{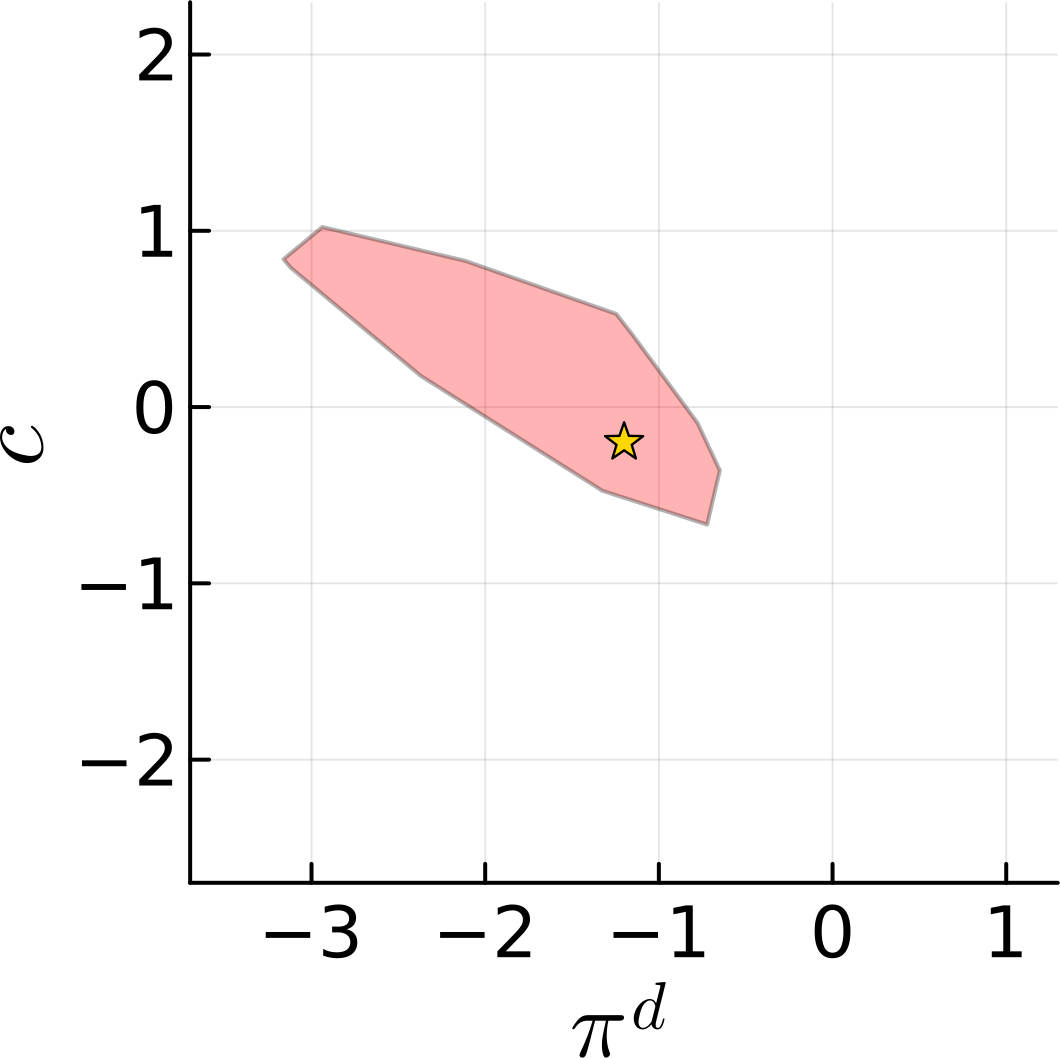}
     \end{subfigure}
     \hfill
     \begin{subfigure}{0.32\textwidth}
         \centering
         \includegraphics[width=\textwidth]{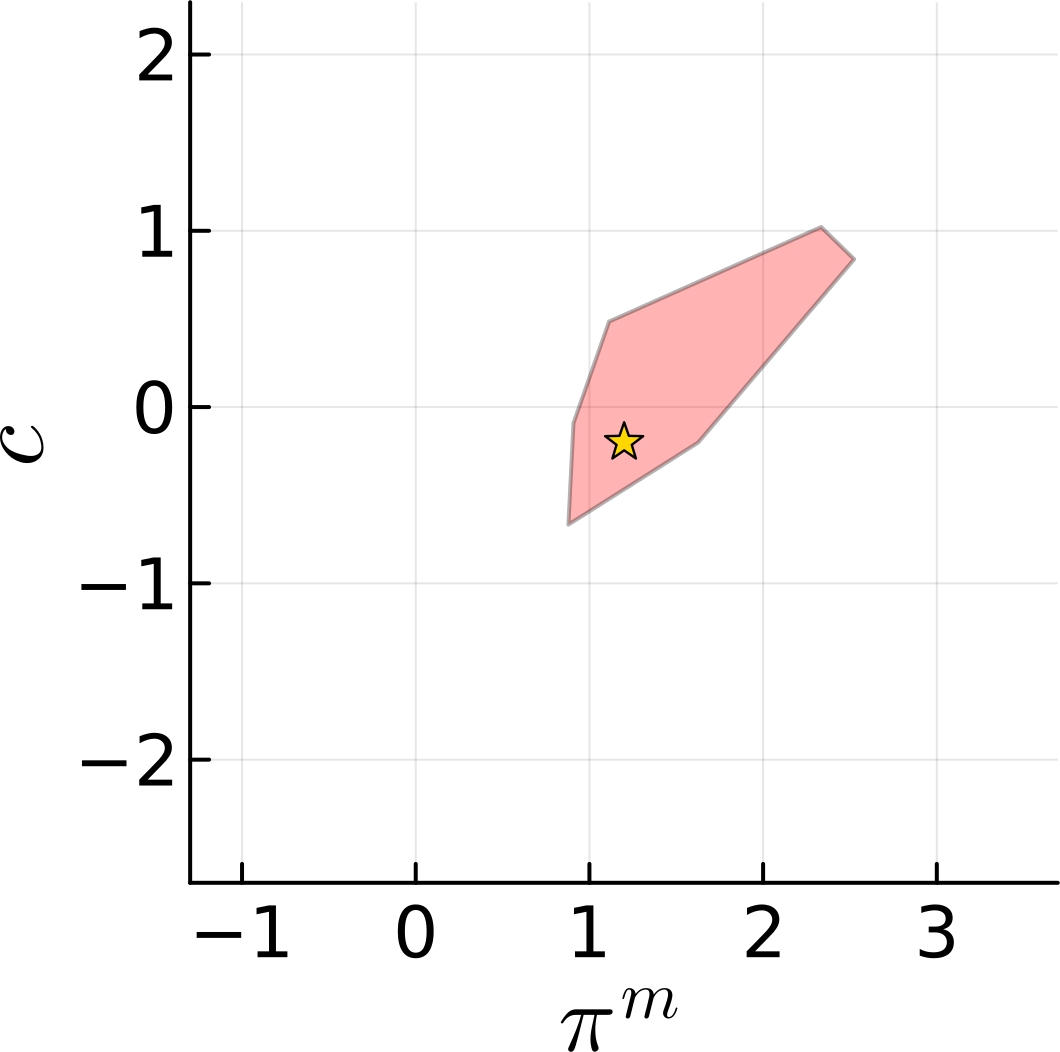}
     \end{subfigure}
     
    \caption{Convex hull of projected identified set}
    \label{fig:PSDprojection}
\end{figure}

Figure \ref{fig:PSDprojection} visualizes the results by plotting the convex hull of the sharp identified set corresponding to equilibrium (ii). In each subfigure, the shaded area represents the convex hull of the projected identified sets, and the star represents the location of the true parameters. As expected, the identified set is non-singleton and contains the true parameter. 

Our numerical example shows that a strong assumption on information is far from innocuous: removing arbitrary assumptions on who observes what may result in a significant loss of identifying power. Robustness to misspecification on players' information structure expands the identified set, but the degree to which the identified set expands can vary depending on which equilibrium is selected in the data generating process.

\section{Estimation} \label{sec:estimation}

We propose computationally feasible estimation strategies by formulating a mathematical program with equilibrium constraints \`{a} la \cite{su2012constrained}. We assume the econometrician knows the true conditional choice probability vector $\phi$ (we relax this assumption in the next section). We also assume that the baseline information structure $S$ is fixed and denote the sharp identified set as $\Theta_I \equiv \Theta_I^{MCE}(S)$.

For a candidate parameter $\theta$, we have $\theta \in \Theta_I$ if and only if there exists a decision rule $\sigma$ that satisfies the Markov correlated equilibrium obedience conditions and induces the observed conditional choice probabilities. We characterize the Markov correlated equilibrium as a set of constraints that the econometrician can plug into optimization software.

\begin{theorem}[Sharp Identified Set] \label{thm:MPEC}
Suppose Assumption \ref{assu:baseline} holds. Then $\theta \in \Theta_I$ if and only if there exists a pair $(\sigma,V)$ that satisfies 
\begin{align}
    &\sigma_{a \vert x, \varepsilon, \tau} \geq 0 \text{ for each $a$ and } \sum_a \sigma_{a\vert x, \varepsilon, \tau} = 1, &&\quad \forall x,\varepsilon,\tau \label{eqn:decision.rule} \\
    & \phi_{a\vert x} = \sum_{\varepsilon, \tau} \psi_{\varepsilon \vert x }\pi_{\tau \vert x, \varepsilon}  \sigma_{a \vert x, \varepsilon, \tau}, &&\quad \forall a,x \label{eqn:consistency.1} \\
    &\sum_{\varepsilon, \tau_{-i}, a_{-i}} \psi_{\varepsilon | x} \pi_{\tau \vert x, \varepsilon} \sigma_{a \vert x, \varepsilon, \tau} \partial v_i^\theta(a_i',a,x,\varepsilon_i) \leq 0, &&\quad \forall i, x, \tau_i, a_i, a_i'  \label{eqn:obedience.1} \\
    & V_{i,x} = \sum_{\varepsilon,\tau,a} \psi_{\varepsilon\vert x} \pi_{\tau\vert x, \varepsilon} \sigma_{a\vert x,\varepsilon,\tau } u_i^\theta (a,x,\varepsilon_i) + \delta \sum_{a,x'} \phi_{a \vert x} V_{i,x'} f_{x'\vert a,x}, &&\quad \forall i,x \label{eqn:value.2}
\end{align}
where $v_i^\theta(a,x,\varepsilon_i) \equiv u_i^\theta(a,x,\varepsilon_i) + \delta \sum_{x'} V_{i,x'} f_{x'\vert a, x}$ is the outcome-specific value function, and $\partial v_i^\theta(a_i',a,x,\varepsilon_i) \equiv v_i^\theta(a_i',a_{-i},x,\varepsilon_i) - v_i^\theta(a_i,a_{-i},x,\varepsilon_i)$ represents the deviation payoff from $a_i$ to $a_i'$.
\end{theorem}
Constraints \eqref{eqn:decision.rule} mean that $\sigma$ is a proper conditional probability distribution, as required by the definition of a decision rule; \eqref{eqn:consistency.1} requires that $\sigma$ induces the observed conditional choice probabilities $\phi$; \eqref{eqn:obedience.1} is the Markov correlated equilibrium obedience condition; \eqref{eqn:value.2} is the ex-ante value function equation described in \eqref{eqn:ex.ante.value.function}.\footnote{Note \eqref{eqn:value.2} also allows us to express each $V_i$ as a closed-form function of $\sigma$. This is similar to expressing the ex-ante value function as a closed-form function of conditional choice probabilities in \citet{Aguirregabiria2007}.} Note that $\psi$, $\pi$, $\phi$, $\delta$, and $f$ are treated as known objects in the mathematical program.\footnote{These conditions bear some resemblance to those in \citet{abreu1990toward} and \citet{athey2004collusion}. Their approach characterizes the set of continuation payoffs using such conditions, raising the question of whether it could also apply in our setting. However, the Markov structure in our framework imposes a tight restriction between stage-game and continuation payoffs, which renders their approach inapplicable.}


\paragraph{Fully Robust Identified Set} In the special case where the analyst sets the baseline information structure to players having no minimal signals, $S^{null}$, the corresponding identified set---``fully robust identified set''---allows for the underlying information structure to be completely arbitrary. It turns out that determining whether a candidate parameter enters the fully robust identified set is computationally easier because the associated program is linear.

\begin{theorem}[Fully Robust Identified Set] \label{thm:fully.robust.identified.set} Suppose \ref{assu:baseline} holds. $\theta \in \Theta_I^{MCE}(S^{null})$ if and only if there exists $(\sigma,V)$ that satisfies the following linear constraints
\begin{align}
    &\sigma_{a\vert x,\varepsilon} \geq 0 \text{ for each $a$ and } \sum_{a} \sigma_{a \vert x, \varepsilon} = 1, && \quad \forall x, \varepsilon \nonumber \\
    &\phi_{a \vert x} = \sum_{\varepsilon} \psi_{\varepsilon \vert x} \sigma_{a \vert x, \varepsilon}, && \quad  \forall a, x \label{eqn:fully.robust.consistency} \\
    & \sum_{\varepsilon,a_{-i}} \psi_{\varepsilon \vert x} \sigma_{a \vert x, \varepsilon} \partial u_i^\theta(a_i',a,x,\varepsilon_i) + \delta \sum_{a_{-i},x'} \phi_{a\vert x} V_{i,x'} \partial f_{x' \vert a_i',a,x} \leq 0, && \quad \forall i, x, a_i, a_i' \label{eqn:fully.robust.obedience.2} \\
    &V_{i,x} = \sum_{\varepsilon,a} \psi_{\varepsilon \vert x} \sigma_{a \vert x, \varepsilon}  u_i^\theta(a,x,\varepsilon_i) + \delta \sum_{a, x'} \phi_{a \vert x} V_{i,x' } f_{x'\vert a,x} , && \quad \forall i, x  \label{eqn:fully.robust.value} 
\end{align}
where $\partial u_i^\theta(a_i',a,x,\varepsilon_i) \equiv u_i^\theta(a_i',a_{-i},x,\varepsilon_i) - u_i^\theta(a,x,\varepsilon_i)$ represents the deviation flow payoff from $a_i$ to $a_i'$, and $\partial f_{x' \vert a_i', a, x} \equiv f_{x' \vert a_i', a_{-i}, x} - f_{x' \vert a, x}$ is the change of the transition probability from $a_i$ to $a_i'$. 
\end{theorem}

The expressions above reflect the assumption that the players receive no signals about the latent state $\varepsilon$. They only rely on their common prior to update their beliefs after observing private recommendations by the information designer. Since $\Theta_I^{MCE}(S) \subseteq \Theta_I^{MCE} (S^{null})$ for any information structure $S$ by Theorem \ref{thm:tighter_set},  the analyst can exploit computational tractability by first estimating the fully robust identified set and then analyzing cases with minimal information specified, as the search area can be restricted. However, the fully robust identified set may be quite large and empirically uninformative \citep[c.f.,][]{Magnolfi2023,koh2023stable}.

\section{Inference} \label{sec:inference}

The inference problem in this paper is characterized by a moment inequality model with a large number of moment inequalities and a high-dimensional nuisance parameter, making it difficult to apply many inference methods for partially identified models.\footnote{See \citet{canay2017practical} and \citet*{canay2023user} for recent surveys of the inference methods for partially identified models.} Inference methods used by \citet*{Syrgkanis2021}, \citet{Magnolfi2023}, and \citet{Gualdani2024} (e.g., \citet*{Chernozhukov2007}'s subsampling approach) are computationally challenging to implement in our setting because the Markov correlated equilibrium constraints are non-convex.

We propose a computationally attractive strategy for inference based on the key insights from \citet{HorowitzLee2022}.\footnote{\citet{koh2023stable} uses a similar approach in static discrete games settings.} Our approach is intended to minimize the need for repeatedly solving non-convex optimization problems. For simplicity, let us assume that the transition probability function is known to the researcher. However, an extension to the case where $f$ needs to be estimated is straightforward. Our approach is also applicable to Bayes correlated equilibrium.

In our setting, the sampling errors are only associated with the reduced-form parameters, namely the conditional choice probabilities (CCPs) $\phi$. If $\phi$ is known, the identified set can be estimated without statistical uncertainty. Thus, we propose controlling for the sampling uncertainty using a confidence set for the CCPs.\footnote{\citet{HorowitzLee2022} proposes methods for carrying out (non-asymptotic) inference when the partially identified parameters are characterized by constraints that involve unknown population means that can be estimated from data. Our approach follows their insights in treating the CCPs as unknown parameters that determine the partially identified set. A similar insight has been employed by \citet{kline2016bayesian}, who propose a Bayesian approach.} 

\subsection{Confidence Set}

Suppose $\Phi_\alpha$ is a confidence set for the CCPs $\phi$ with the property that $\phi \in \Phi_\alpha $ with probability at least $1-\alpha$ as the sample size $n$ goes to infinity. Leading examples of $\Phi_\alpha$ are box constraints or ellipsoids. Let $\Theta_I(\phi)$ be the identified set when the CCPs are $\phi$. Define the confidence set as
\begin{equation*}
    \widehat{\Theta}_I^\alpha \equiv \bigcup_{\phi \in \Phi_\alpha} \Theta_I(\phi). 
\end{equation*}

If $\Phi_\alpha$ covers the true $\phi$ with high probability, then $\widehat{\Theta}_I^\alpha$ covers the true $\Theta_I$ with high probability. Furthermore, checking whether $\theta \in \widehat{\Theta}_I^\alpha$ amounts to solving a nonlinear program in Theorem \ref{thm:MPEC}, but with $\phi$ as part of the optimization variables subject to the constraint $\phi \in \Phi_\alpha $.
\begin{theorem}\label{thm:confidence.set}
    Let $\Phi_\alpha$ be a confidence region for $\phi$ such that
    \begin{equation*}
        \underset{n \to \infty}{\lim \inf} \mathrm{Pr} \left( \phi \in \Phi_\alpha \right) \geq 1-\alpha.
    \end{equation*}
    Then 
    \begin{equation*}
        \underset{n \to \infty}{\lim \inf} \mathrm{Pr} \left( \Theta_I \subseteq \widehat{\Theta}_I^\alpha \right) \geq 1-\alpha.
    \end{equation*}
    Moreover, $\theta \in \widehat{\Theta}_I^\alpha$ if and only if there exists $(\sigma,V,\phi)$ subject to $\phi \in \Phi_\alpha$, \eqref{eqn:decision.rule}, \eqref{eqn:consistency.1}, \eqref{eqn:obedience.1}, and \eqref{eqn:value.2}.
\end{theorem}

\subsection{Implementation}
We briefly discuss possible approaches for constructing $\Phi_\alpha$. While there are many possible approaches available, it is helpful to  construct $\Phi_\alpha$ as linear constraints because they are computationally easy to handle. For example, one can construct simultaneous confidence intervals of the form
\begin{equation*}
    \Phi \equiv \left \{ \phi: \hat{L}_{a\vert x} \leq \phi_{a\vert x} \leq \hat{U}_{a\vert x}, \quad \forall a, x \right\}
\end{equation*}
where the vectors $\hat{L}, \hat{U} \in \mathbb{R}^{\vert \mathcal{A} \vert \times \vert \mathcal{X} \vert}$ are determined by data. Examples include simultaneous confidence intervals for multinomial probabilities proposed by \cite{fitzpatrick1987quick} or sup-$t$ band by \cite{montiel2019simultaneous}; the former is easier to compute since the endpoints of the intervals can be computed in closed-form, but the latter can be tighter. 

An alternative is to take a parametric approach.\footnote{With limited observations relative to covariate dimension, nonparametric CCP estimation may fail, as some cells contain no data. A flexible multinomial logit specification, though inconsistent under misspecification, offers lower variance and often outperforms nonparametric estimators in small samples. \cite{Aguirregabiria2007} show this approach performs well in a two-step pseudo–maximum likelihood framework.} Suppose $\phi = \phi(\gamma)$, where $\gamma$ is the parameter vector that governs the CCPs. One may estimate $\gamma$ and construct simultaneous confidence intervals of the form
\begin{equation*}
    \Phi \equiv \left \{ \gamma: \hat{L}_k \leq \gamma_k \leq \hat{U}_k, \quad \forall k \right \}
\end{equation*}
where $\hat{L},\hat{U} \in \mathbb{R}^{\dim(\gamma)}$ can be computed using the asymptotic distribution of $\hat{\gamma}$. One may also consider replacing $\phi$ with its linear approximation $\phi(\gamma) \approx \phi(\hat{\gamma}) + \nabla \phi(\hat{\gamma})^\top (\gamma - \hat{\gamma})$ to further simplify the computation. In the identified set search, we also add the constraints defining $\phi$ as proper conditional distributions, i.e., $\phi_{a\vert x} \geq 0$, $\forall a,x$, and $\sum_{a} \phi_{a\vert x} = 1$, $\forall x$.

In our empirical application, we construct simultaneous confidence intervals using the sup-$t$ band approach, following \citet{montiel2019simultaneous}.

\section{Empirical Application} \label{sec:empirical.application}

We study a dynamic entry game by the two largest players in the US coffee chain industry, Starbucks and Dunkin', to illustrate the usefulness of our framework. The US coffee chain industry has grown significantly in recent decades, driven by increasing demand for specialty coffee and the rise of a culture emphasizing quality, convenience, and socialization. While companies compete for market share through pricing, product differentiation, marketing strategies, and in-store service, this market is highly concentrated. In particular, Starbucks and Dunkin' have significantly outpaced other competitors by a wide margin. Starbucks and Dunkin' accounted for 65\% and 28\% (resp. 55\% and 35\%) of total sales (resp. outlets) from the top 8 coffee chains in the US in 2019 \citep{technomic2019}.\footnote{The third and fourth chains were Tim Hortons and Dutch Bros., which accounted for 2.2\% and 1.7\% (resp. 2.6\% and 1.3\%) of the sales (resp. outlets) out of the top 8 chains.}

Little is publicly known about how coffee chains predict opponents' strategic entry decisions. It is reasonable to expect that firms have good information about their stores' (potential) profitability. Still, less is known about how firms learn about their opponents' profitability. Resorting to the standard ``incomplete'' information assumption---firms observe their payoff shocks but observe nothing about their opponents' shocks---will likely result in a model misspecification. Whether this misspecification is an empirical concern depends on how much it may change the structural estimates and counterfactual outcomes. We apply our Markov correlated equilibrium framework to estimate the parameters and conduct a counterfactual experiment with weaker assumptions on players' information.

\subsection{Data and Descriptive Evidence}
\begin{figure}[htb!]
    \centering
        \includegraphics[width=0.7\textwidth]{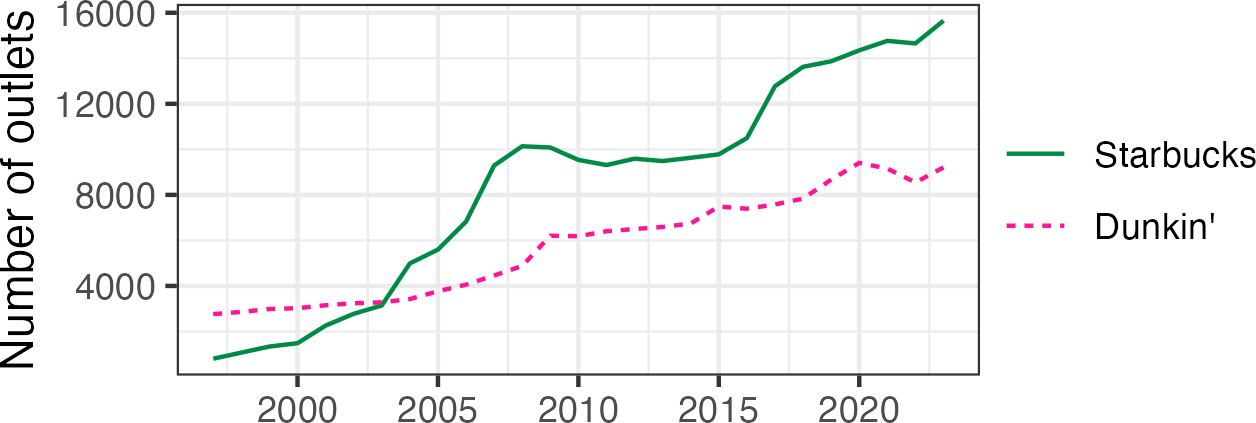}
    \caption{Number of Starbucks and Dunkin' stores in the US over time}
    \label{fig:store.counts}
\end{figure}

\begin{figure}[htb!]
\centering
  \includegraphics[width=\textwidth]{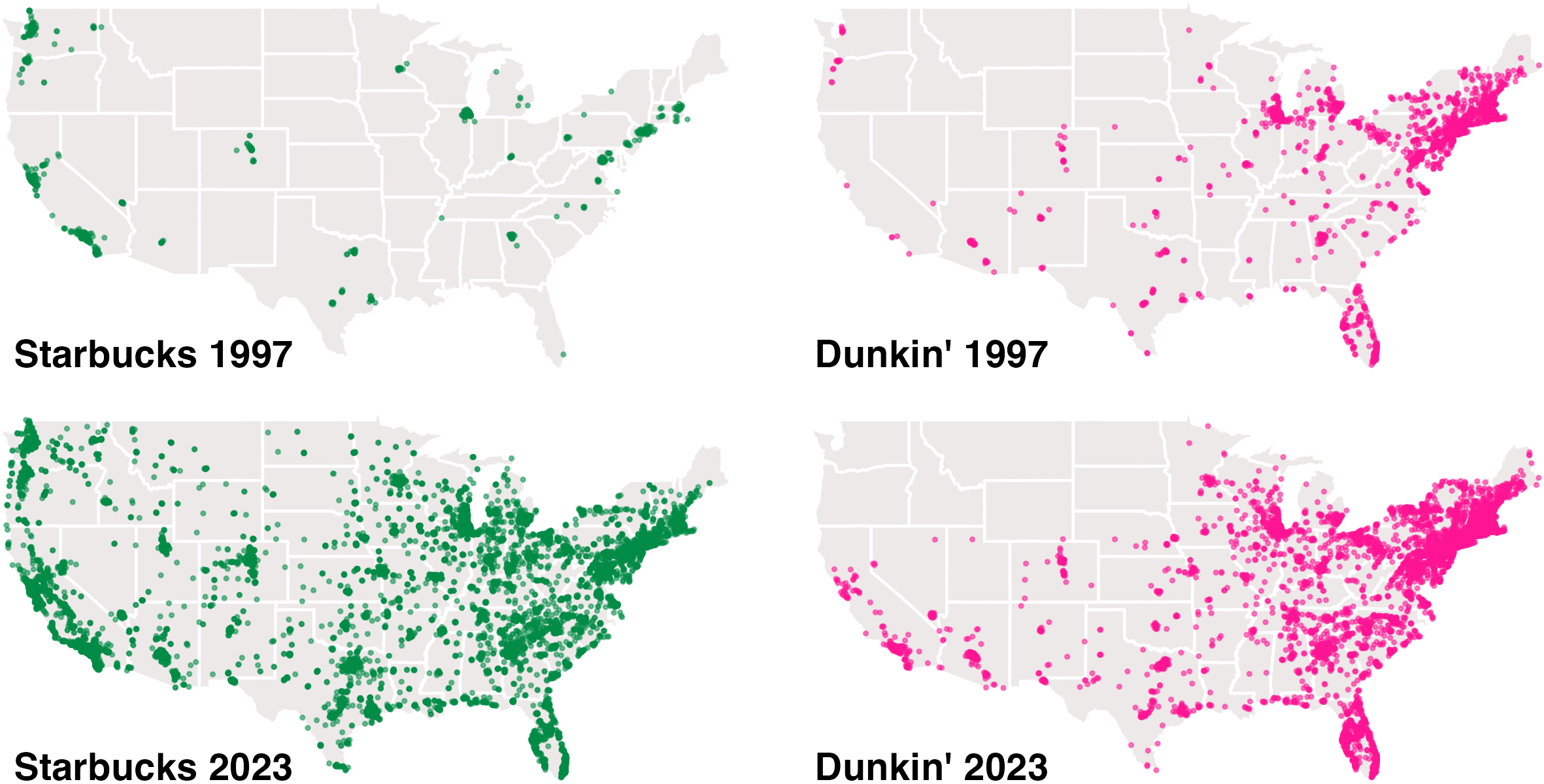}
\caption{Growth of Starbucks and Dunkin'}
\label{fig:growth.of.starbucks.and.dunkin}
\end{figure}

Our primary dataset is Data Axle's U.S. Historic Business Database, which has annual location information of all business establishments in the U.S. from 1997 to 2023. Figure \ref{fig:store.counts} shows that the number of Starbucks and Dunkin' stores in the US increased dramatically from 1997 to 2023. Figure \ref{fig:growth.of.starbucks.and.dunkin} illustrates how the firms dramatically expanded their geographic footprint over time.

We define markets as four-digit 2010 census tracts.\footnote{Four-digit census tracts are census-defined geographic units that are contiguous and strike a balance between county (which is too large) and six-digit census tracts (which is too small).} Our geographic market definition assumes that coffee chains compete in small geographic areas.\footnote{\citet{guler2018inferring} shows that demand for Starbucks is highly localized. \citet{thomadsen2005effect} shows evidence that only fast-food chains within approximately 0.5 miles---which amounts to 10 minutes walking distance---compete as close substitutes in California; it is likely that coffee consumers travel even less.} A firm is active in a market when we observe the firm's outlet in the market. We supplement the firm-location data with the demographic/geographic data from the Longitudinal Track Data Base \citep*{logan2014interpolating} and the neighborhood characteristics data from the National Neighborhood Data Archive. We limit our sample period to 2003--2017 to match the sample period of the neighborhood characteristics data.\footnote{We still use the business location data from 2018 to reduce the possibility of incorrectly determining firms not being active due to missing data. We consider a firm to have switched from active to not active in a market (i.e., exited the market) when we do not observe the firm's outlet in the market for the following five years.} We exclude the markets without any eating places during the sample period.

Table \ref{tab:active.probabilities.own.outlets.county} Panel A shows the active probabilities of each firm from 2018--2023 when the firm either did not have any outlets or had at least one outlet in the corresponding county from 2003--2017. The {\it ex-post} active probability in a county is much lower when the firm does not have any outlets in that county compared to when it does (1\% vs. 20\%). Thus, it is reasonable to consider that a firm does not take into account the competitor's situation if the competitor does not have any outlets in the county, and we define a firm as a potential entrant in each market if the firm had at least one outlet in the corresponding county during the sample period. We also use own-chain outlets in the county as firm-specific excluded variables that capture the network effects associated with economies of density \citep{guler2018inferring}. Table \ref{tab:active.probabilities.own.outlets.county} Panel B demonstrates that the number of own-chain outlets in the county has a positive effect on the active probability in markets where both Starbucks and Dunkin' are potential entrants (two-player markets). In contrast, the number of competitor-chain outlets in the county affects the active probability little. We interpret the weak negative relationship as stemming from the firm's strategic behavior in response to the higher competitor's active probability.

\begin{table}[htb!]

\caption{\label{tab:active.probabilities.own.outlets.county}Active probabilities by the number of outlets in county}
\centering
\begin{tabular}[t]{lcc}
\toprule
 & Starbucks & Dunkin'\\
\midrule
A. Active probability after 2017 &  & \\
\hspace{1em}Own-chain outlets in county $= 0$ & $0.011$ & $0.011$\\
\hspace{1em}Own-chain outlets in county $> 0$ & $0.211$ & $0.197$\\
\addlinespace
B. Active probability in two-player markets &  & \\
\hspace{1em}Own-chain outlets in county $\leq$ median & $0.095$ & $0.043$\\
\hspace{1em}Own-chain outlets in county $>$ median & $0.195$ & $0.234$\\
\hspace{1em}Competitor-chain outlets in county $\leq$ median & $0.163$ & $0.146$\\
\hspace{1em}Competitor-chain outlets in county $>$ median & $0.126$ & $0.130$\\
\bottomrule
\end{tabular}
\end{table}

Table \ref{tab:summary.statistics.of.coffee.chain.data} reports the summary statistics of a balanced panel of $M=38,687$ markets observed over $T = 15$ years. Both Starbucks and Dunkin' are potential entrants in 79\% of the markets, and 17\% and 4\% respectively have Starbucks and Dunkin' as a single potential entrant. The firms are active in 14\% and 11\% of the observations, while entries and exits are rare (1\% and 0.2\%, respectively), indicating a high correlation in the incumbency status. We assume the firms take the market characteristics as exogenous and time-invariant. For each market, we average characteristics over the observation period and treat them as fixed attributes. The key common characteristic is the number of eating places, which captures overall market attractiveness, as customer traffic tends to rise with the density of food establishments.

\begin{table}[htb!]

\caption{\label{tab:summary.statistics.of.coffee.chain.data}Summary statistics of 38,687 coffee chain markets from 2003 to 2017}
\centering
\begin{tabular}[t]{lccc}
\toprule
A. Potential entrant & Both & Starbucks & Dunkin'\\
\hspace{1em}Market share & $0.786$ & $0.172$ & $0.042$\\
\addlinespace
B. Player decisions & Starbucks & Dunkin' & \\
\hspace{1em}Active probability & $0.139$ & $0.113$ & \\
\hspace{1em}Entry probability & $0.011$ & $0.007$ & \\
\hspace{1em}Exit probability & $0.002$ & $0.002$ & \\
\addlinespace
C. Market characteristics & Mean & Std. Dev. & Range\\
\hspace{1em}Number of eating places & $15.6$ & $25.2$ & $[0.1, 649.8]$\\
\hspace{1em}Starbucks outlets in county & $592.0$ & $1243.1$ & $[0.0, 6423.0]$\\
\hspace{1em}Dunkin' outlets in county & $354.9$ & $750.9$ & $[0.0, 3948.0]$\\
\bottomrule
\end{tabular}
\end{table}

\subsection{Econometric Model}

Our model follows the standard dynamic oligopoly entry/exit model with heterogeneous firms \citep*{Aguirregabiria2007,Pakes2007,pesendorfer2008asymptotic}. In each market $m$ and time period $t$, firm $i$ decides whether to operate in the market ($a_{imt}=1$) or not ($a_{imt}=0$). An active firm's flow profit is
\begin{equation} \label{eqn:empirical.model}
    u_i^\theta(a_{imt}=1, a_{jmt},x_{mt},\varepsilon_{imt}) = \theta_{w}^\top w_{m} + \theta_{i,v} v_{im} + \theta_{ec}(1 - z_{imt}) + \theta_{i,ce} a_{jmt} + \varepsilon_{imt}.
\end{equation}
Here, public state $x_{mt}$ includes three components, common time-invariant market characteristics $w_{m}$, firm-specific time-invariant excluded variables $(v_{im}, v_{jm})$, and incumbency status $(z_{imt} \equiv a_{im,t-1}, z_{jmt} \equiv a_{jm,t-1})$. $w_{m}$ includes a constant and log number of eating places, and $v_{im}$ is log number of own-chain outlets in the county. Parameters $\theta_{w}$, $\theta_{i,v}$, $\theta_{ec}$, and $\theta_{i,ce}$ represent coefficients on market characteristics, excluded variable, entry cost, and competitive effects, respectively. We allow the coefficients on the excluded variable and competitive effects to be asymmetric. $\varepsilon_{imt}$ is the idiosyncratic per-period payoff shock, which is assumed to be independently drawn from the standard logistic distribution. The state transition is deterministic since the only dynamic state variable is the incumbency status, determined by the firms' actions in the previous period. We set the discount factor to $\delta = 0.9$ and the flow payoff from being inactive to zero.\footnote{Normalizing the flow profit from the outside option to zero is not without loss of generality for counterfactual analysis \citep*{aguirregabiria2014identification, kalouptsidi2017non, kalouptsidi2021identification}. We show that our counterfactual experiment of reducing a proportion of entry cost is equivalent to providing an entry subsidy equal to the same proportion of the entry cost net of scrap value without this normalization in Online Appendix \ref{sec:interpretation.of.counterfactual.analysis}. } We discretize market characteristics $(w_{m},v_{im})$ into two bins by grouping the observations above and below the median and replacing the values with within-bin means.

We first estimate the parameters under the standard incomplete information Markov perfect equilibrium assumption using the $K$-pseudo maximum likelihood approach \citep{Aguirregabiria2007, bugni2021iterated}. Each player observes their own idiosyncratic shock $\varepsilon_{imt}$ but not the opponent's shock $\varepsilon_{jmt}$. We then estimate the Markov correlated equilibrium identified set, maintaining this standard informational assumption. The identified set is informationally robust in the sense that it captures all possible parameters that could generate the data under Markov perfect equilibrium, allowing for the possibility that the players know their own $\varepsilon_{imt}$ and may observe information about the opponent's $\varepsilon_{jmt}$ before making decisions. We also estimate the fully robust identified set, which additionally encompasses the possibility that the players do not observe or are partially informed about their $\varepsilon_{imt}$ (formally defined in Section \ref{sec:estimation}). The identified set is fully robust to any misspecification of the information structure regarding the idiosyncratic shocks $\varepsilon_{mt} = (\varepsilon_{imt},\varepsilon_{jmt}) $. Lastly, we estimate the model without the excluded variable $v_{im}$ with these three information structures above to gauge the importance of the exclusion restriction. We outline our estimation strategy of the Markov correlated equilibrium identified sets and relegate the computational details to Online Appendix \ref{sec:computation.estimation}.

We leverage the excluded variable with large support as discussed in Section \ref{sec:properties.informationally.robust.identified.set}. Our excluded variable $v_{im}$, the log number of own-chain outlets in the county, has large support because it has no lower bound.\footnote{Given the length of our observation period, we assume that this variable sufficiently captures the time-persistent match effect between firms and counties. While our estimation procedure relies on a specific parametric form of $v_{im}$, sufficient variation in the excluded variable will tighten the identified set even if we impose another parametric form (see e.g., \citet{Magnolfi2023}).} Firm $i$ is not a potential entrant (i.e., $i$'s active probability is $0$) if the number of own-chain outlets in the county is $0$, since $v_{im} = \log 0 = -\infty$. This model specification is motivated by the fact that a firm will be rarely active in markets where the firm did not have any outlets in the relevant county (Table \ref{tab:active.probabilities.own.outlets.county} Panel A). In markets where firm $j$ is not a potential entrant or $a_{jmt}$ is always $0$, firm $i$ solves a single-agent dynamic decision problem, as the opponent's idiosyncratic shock $\varepsilon_{jmt}$ becomes payoff-irrelevant. We estimate the single-agent discrete choice model, which contains the payoff parameters except the competitive effects parameters in \eqref{eqn:empirical.model},\footnote{We can identify the competitive effects parameters if there is a subset of markets defined by excluded variables where the opponent is always active additionally. It would be challenging to find a sufficient number of such markets to reasonably estimate the competitive effects parameters in our setting (see Table \ref{tab:active.probabilities.own.outlets.county} Panel B for the effect of our excluded variable on the active probability).} using the two-step pseudo-likelihood method \citep{Aguirregabiria2002}. This step allows us to obtain estimates for non-competitive effects parameters, $\theta_w, \theta_{i,v}, \theta_{j,v}$, and $\theta_{ec}$.

We then apply Theorem \ref{thm:confidence.set} to estimate the identified set of the competitive effects parameters using markets where both firms are potential entrants. We adopt the non-competitive effects parameters estimated from the single-agent model in the Markov correlated equilibrium identified set estimation to demonstrate the identifying power of our Markov correlated equilibrium framework.\footnote{\citet{aradillas2022inference} also takes a similar two-step approach in static ordered-response games. The authors point identify the non-competitive effects parameters and use the estimates to partially identify the competitive effects parameters.} We estimate the CCPs using flexible logit specification and construct their simultaneous confidence intervals with a plug-in sup-$t$ implementation \citep{montiel2019simultaneous}. We estimate a criterion function $Q(\theta)$ that measures the degree of violation in the obedience condition \eqref{eqn:obedience.1} for given competitive effects parameters $\theta = (\theta_{i,ce},\theta_{j,ce})$. We construct identified set estimates by grid searching $\theta$ that satisfies the estimated $Q(\theta) \leq c$ for a small threshold $c > 0$. We use non-zero $c$ to accommodate misspecifications in the structural and CCP logit models. Larger values of $c$ yield more conservative identified sets. We do not incorporate the sampling error associated with the first-stage single-agent model estimation.

We also consider a model without the excluded variables $v_{im}$ and $v_{jm}$. We grid search for all the payoff parameters $\theta = (\theta_{w},\theta_{ec},\theta_{i,ce},\theta_{j,ce}) $ for this model.

\subsection{Estimation Results}

Table \ref{tab:estimation.results} reports our estimation results with the exclusion restriction. The presence of an opponent has a positive spillover effect on profitability under the standard incomplete information Markov perfect equilibrium assumption (Column (1)). The 95\% confidence intervals (CIs) of the competitive effects parameters become wider as we make the information structure more flexible (compare columns from left to right).\footnote{We set the criterion function threshold $c$ so that the CIs of the competitive effects parameters become column (1) $\subset$ column (2) $\subset$ column (3).} The Markov correlated equilibrium estimates suggest the possibility of a large negative opponent effect (columns (2) and (3)). However, the positive effect of Starbucks' presence on Dunkin's profitability is tightly bounded above. The exclusion restriction helps narrow the direction of Dunkin's competitive effects parameter (compare with the results without the excluded variable in Table \ref{tab:estimation.results.woexclusion} in Appendix \ref{sec:results.wo.exclusion}).

\begin{table}[htb!]

\caption{\label{tab:estimation.results}Structural parameter estimates}
\centering
\resizebox{\linewidth}{!}{\begin{threeparttable}
\begin{tabular}[t]{lccc}
\toprule
 & \makecell{MPE\\Standard incomplete\\(1)} & \makecell{MCE\\Standard incomplete\\(2)} & \makecell{MCE\\Fully robust\\(3)}\\
\midrule
Intercept & $[-0.032, -0.002]$ & $[-0.040, 0.040]$ & $[-0.040, 0.040]$\\
 & $-0.016$ & $0.000$ & $0.000$\\
Eating places (log) & $[0.103, 0.110]$ & $[0.117, 0.136]$ & $[0.117, 0.136]$\\
 & $0.106$ & $0.127$ & $0.127$\\
Starbucks outlets in county (log) & $[0.027, 0.031]$ & $[0.019, 0.029]$ & $[0.019, 0.029]$\\
 & $0.029$ & $0.024$ & $0.024$\\
Dunkin' outlets in county (log) & $[0.027, 0.031]$ & $[0.029, 0.048]$ & $[0.029, 0.048]$\\
 & $0.029$ & $0.038$ & $0.038$\\
Entry cost & $[-8.473, -8.365]$ & $[-8.323, -8.008]$ & $[-8.323, -8.008]$\\
 & $-8.408$ & $-8.165$ & $-8.165$\\
Starbucks competitive effect & $[0.027, 0.049]$ & $[-0.123, 0.143]$ & $[-0.257, 0.449]$\\
 & $0.040$ & - & -\\
Dunkin' competitive effect & $[0.011, 0.035]$ & $[-0.299, 0.038]$ & $[-0.300, 0.074]$\\
 & $0.024$ & - & -\\
\bottomrule
\end{tabular}
\begin{tablenotes}[para]
\footnotesize
\item \textit{Note:} Column (1) reports the 95\% confidence intervals (CIs) and point estimates for the model with the standard incomplete information Markov perfect equilibrium (MPE) assumption. Columns (2) and (3) use our Markov correlated equilibrium (MCE) framework with the standard incomplete information and without minimal signals, respectively. The MCE column parameters except the competitive effects parameters are obtained using single-agent discrete choice estimation using data where the firms operate as single players. CIs are bootstrapped with resampling at the market level. The competitive effects parameters estimated from the MCE framework do not account for the sampling error associated with the first-stage single-agent dynamic discrete choice estimation.
\end{tablenotes}
\end{threeparttable}}
\end{table}

The market characteristics coefficients and entry cost from the two-player model (column (1)) and the single-agent model (columns (2) and (3) display the same estimates) are similar, which justifies the usage of different subsets of markets to estimate the same parameters.\footnote{The CIs do not overlap for some parameters. The discrepancies can be attributed to either misspecification in the information structure or the differences in the parameters in different markets. We do not provide a formal test of misspecification in the information structure. See \citet*{han2024testing} for a related test in the Bayes correlated equilibrium context.} The market characteristics coefficient estimates show positive effects of attractive markets and economies of density on profitability. The entry cost estimates are relatively large compared to the other parameter estimates, which is consistent with the low entry probabilities.

\subsection{Counterfactual}
\label{sec:counterfactual}
As a counterfactual experiment, we reduce each firm's entry cost by 10\% and simulate the number of active firms from 2003 under three assumptions on information structures used in the estimation. The first is the standard incomplete information Markov perfect equilibrium predictions. We then apply the Markov correlated equilibrium to make informationally robust predictions, assuming either the standard incomplete information or the fully robust specification. For each case, we use the structural parameters estimated under the corresponding solution concept and information structure and apply the same framework to compute the counterfactual predictions.\footnote{In the counterfactual Markov correlated equilibrium predictions, we do not use the \emph{latent information structures} that generate the Markov perfect equilibrium predictions compatible with the data for each parameter value in the Markov correlated equilibrium identified set estimates. \citet*{Bergemann2022} shows how to conduct counterfactual analysis with the latent information structure fixed at the pre-counterfactual Markov perfect equilibrium. This approach would not tighten our standard incomplete information Markov correlated equilibrium predictions since the predictions for a given set of structural parameters are tight (Table \ref{tab:flexible.information.structure.counterfactual}). It may appreciably narrow the fully robust predictions by restricting the range of predictions for a given set of structural parameters.}

We consider two extreme scenarios, low and high, to get a sense of a reasonably possible range of the counterfactual predictions. For the low (high) scenario, we use the lower (upper) bounds of the 95\% CIs of the competitive effects parameters and find an equilibrium that minimizes (maximizes) the expected number of active firms.\footnote{Note that the pair of values at the lower bounds of the competitive effects parameters for Starbucks and Dunkin' may not necessarily belong to the confidence set.} As a separate exercise, we also make counterfactual predictions using the lower and upper bounds of the 95\% CIs of all structural parameters, rather than only the competitive effects parameters, to incorporate the uncertainty of the non-competitive effects parameters. We use the counterfactual equilibrium CCPs to simulate the evolution of active firms. We detail the counterfactual equilibrium calculations in Online Appendix \ref{sec:computation.equilibrium}.

\begin{figure}[htb!]
     \centering
     \begin{subfigure}{0.49\textwidth}
         \centering
         \includegraphics[width=\textwidth]{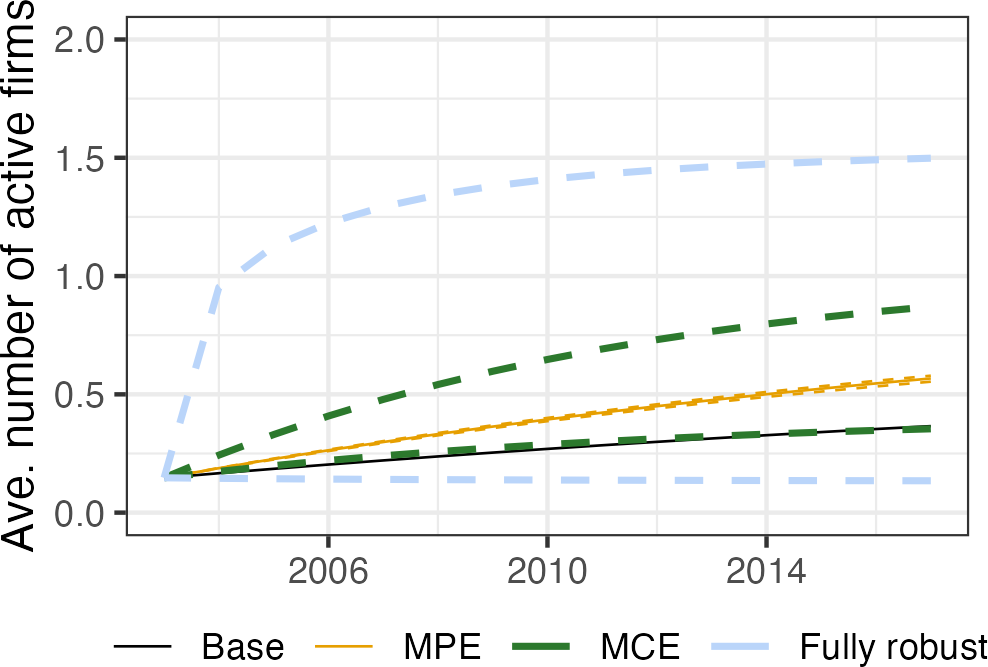}
         \caption{Use CIs of competitive effects}
         \label{fig:counterfactual.ci.competitive}
     \end{subfigure}
     \hfill
     \begin{subfigure}{0.49\textwidth}
         \centering
         \includegraphics[width=\textwidth]{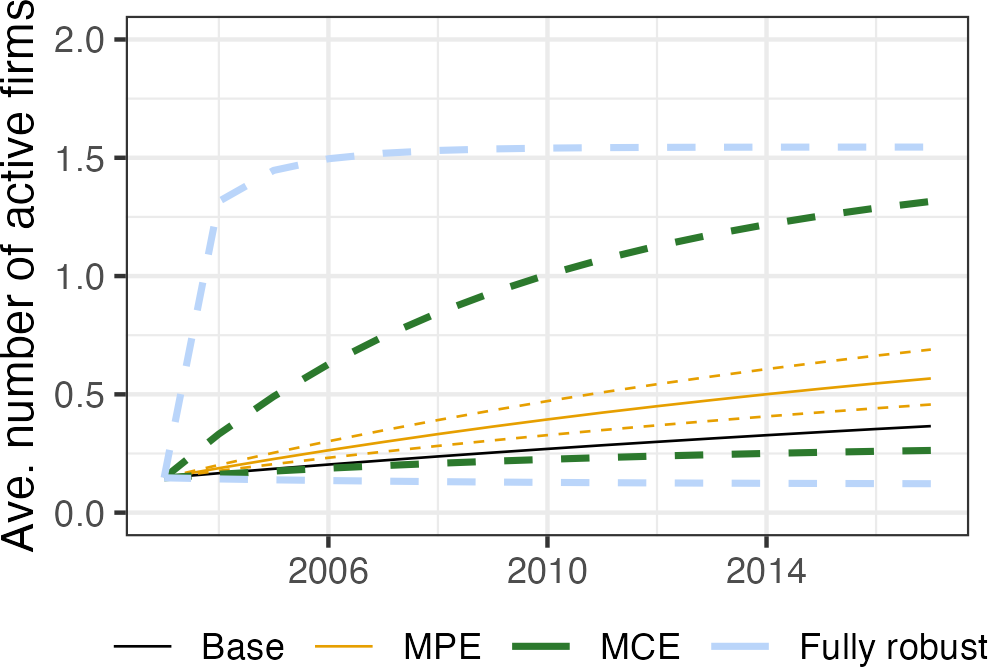}
         \caption{Use CIs of all parameters}
         \label{fig:counterfactual.ci.all}
     \end{subfigure}
    \caption{Counterfactual market status (low/high scenarios in dashed lines)}
    \label{fig:counterfactual}
\end{figure}

Figure \ref{fig:counterfactual} plots the counterfactual market evolution from 2003--2017. The y-axis takes the average of all markets, so the minimum and maximum possible values are 0 (no firms active) and 2 (both firms active), respectively. The no policy baseline with the standard incomplete information Markov perfect equilibrium predicts an increase from $0.15$ to $0.37$ during the sample period (line Base). The Markov perfect equilibrium predicts that the counterfactual policy intervention of reducing the entry cost by 10\% increases the number of active firms by 50\% in 2017 (line MPE). The range of the two scenarios (dashed curves) is very tight using the CIs of the competitive effects parameters (Figure \ref{fig:counterfactual.ci.competitive}, average number of active firms in 2017: low, $0.55$; high, $0.58$) and become somewhat wider using the CIs of all structural parameters (Figure \ref{fig:counterfactual.ci.all}, low, $0.46$; high, $0.69$).

The counterfactual Markov correlated equilibrium predictions, keeping the standard incomplete information, of the average number of active firms in 2017 range from $0.35 $ (low) to $0.87 $ (high) using the CIs of the competitive effects parameters (Figure \ref{fig:counterfactual.ci.competitive}, line MCE). If we are concerned about the informational robustness regarding the competitive effects parameters, this range provides a reasonably plausible set of outcomes, allowing for robust yet still informative policy implications. In contrast, the predictions become much more spread out and less informative, ranging from $0.26$ (low) to $1.32$ (high), using the CIs of all parameters (Figure \ref{fig:counterfactual.ci.all}, line MCE). The fully robust Markov correlated equilibrium generates very wide predictions even right after the start of the counterfactual policy intervention (line Fully robust). Moreover, without the exclusion restriction, the counterfactual Markov correlated equilibrium predictions cover almost the entire range of the average number of active firms, $[0,2]$, even under the standard incomplete information (Figure \ref{fig:counterfactual.wo.exclusion} in Appendix \ref{sec:results.wo.exclusion}).

The counterfactual Markov correlated equilibrium predictions involve two sources of multiplicity: partial identification and estimation uncertainty of structural parameters; and flexible counterfactual information structure. The flexible information structure can generate various counterfactual equilibria, resulting in a wide range of predictions. Table \ref{tab:flexible.information.structure.counterfactual} tabulates the range of counterfactual predictions for a given set of structural parameters to examine the contribution of these multiplicities. The minimum and maximum number of active firms that can arise from the counterfactual Markov perfect equilibrium predictions coincide for a given set of structural parameters (columns under MPE Standard incomplete), which suggest that we do not have multiple equilibria for a fixed vector of parameters and information structure. Thus, the range of Markov perfect equilibrium predictions is entirely driven by the estimation uncertainty (or 95\% CIs) of structural parameters.

\begin{table}[htb!]

\caption{\label{tab:flexible.information.structure.counterfactual}Contribution of flexible information structure in the counterfactual}
\centering
\resizebox{\linewidth}{!}{\begin{threeparttable}
\begin{tabular}[t]{lcccccccc}
\toprule
\multicolumn{1}{c}{} & \multicolumn{1}{c}{} & \multicolumn{3}{c}{\makecell{MPE\\Standard incomplete}} & \multicolumn{2}{c}{\makecell{MCE\\Standard incomplete}} & \multicolumn{2}{c}{\makecell{MCE\\Fully robust}} \\
\cmidrule(l{3pt}r{3pt}){3-5} \cmidrule(l{3pt}r{3pt}){6-7} \cmidrule(l{3pt}r{3pt}){8-9}
Year & Baseline & $\hat{\theta}$ point & $\hat{\theta}$ lower & $\hat{\theta}$ upper & $\hat{\theta}$ lower & $\hat{\theta}$ upper & $\hat{\theta}$ lower & $\hat{\theta}$ upper\\
\midrule
\addlinespace[0.3em]
\multicolumn{9}{l}{A. Use 95\% CIs of competitive effects parameters}\\
\hspace{1em}2003 & $0.15$ & $0.15$ & $0.15$ & $0.15$ & $0.15$ & $0.15$ & $0.15$ & \vphantom{1} $0.15$\\
\hspace{1em}2010 & $0.27$ & $0.39$ & $0.39$ & $0.40$ & $[0.29, 0.32]$ & $[0.64, 0.65]$ & $[0.14, 0.52]$ & $[0.33, 1.41]$\\
\hspace{1em}2017 & $0.37$ & $0.57$ & $0.55$ & $0.58$ & $[0.35, 0.38]$ & $[0.87, 0.87]$ & $[0.13, 0.54]$ & $[0.46, 1.50]$\\
\addlinespace[0.3em]
\multicolumn{9}{l}{B. Use 95\% CIs of all parameters}\\
\hspace{1em}2003 & $0.15$ & $0.15$ & $0.15$ & $0.15$ & $0.15$ & $0.15$ & $0.15$ & $0.15$\\
\hspace{1em}2010 & $0.27$ & $0.39$ & $0.33$ & $0.47$ & $[0.23, 0.23]$ & $[1.00, 1.01]$ & $[0.13, 0.27]$ & $[0.27, 1.54]$\\
\hspace{1em}2017 & $0.37$ & $0.57$ & $0.46$ & $0.69$ & $[0.26, 0.26]$ & $[1.31, 1.32]$ & $[0.12, 0.31]$ & $[0.36, 1.55]$\\
\bottomrule
\end{tabular}
\begin{tablenotes}[para]
\footnotesize
\item \textit{Note:} Values are the simulated average number of active firms. The baseline is the standard incomplete information Markov perfect equilibrium (MPE) predictions without entry subsidy. The other columns show the counterfactual predictions with entry subsidy. Column $\hat{\theta}$ point uses the point estimates of the structural parameters. Panel A and B use the lower/upper bound of 95\% confidence intervals (CIs) of competitive effects parameters and all parameters, respectively, for $\hat{\theta}$ lower/upper. The range of the MCE counterfactuals shows the minimum and maximum predictions for a given set of structural parameters.
\end{tablenotes}
\end{threeparttable}}
\end{table}

The prediction ranges for a given set of structural parameters remain very tight using the Markov correlated equilibrium, using the standard incomplete information as the baseline (columns under MCE Standard incomplete). These prediction ranges encompass the counterfactual outcomes arising from any Markov perfect equilibrium when the players minimally observe their payoff shocks and may observe more information about opponents' payoff shocks. The tight counterfactual outcome bounds indicate that the ability to obtain more information about opponents' payoff shocks and the possibility that such information is correlated across firms have little impact on the outcome, given the model parameters fixed. The small competitive effects parameters (in absolute values) can explain the tight counterfactual bounds. If the competitive effects are null, then the firms do not care about the opponents' behavior, and there is no room for an information designer to affect the total number of entrants, on average.\footnote{The information designer can still induce correlation in the firms’ actions. As a result, the CCPs may deviate from those predicted under the standard MPE scenario.} Here, the competitive effects parameters range between $-0.30$ and $0.14$, which is small considering the standard deviation of the idiosyncratic payoff shock (the standard deviation of the standard logistic random variable is $\sqrt{\pi^2/3} \approx 1.81$). Consequently, the degree to which the counterfactual outcomes can vary depends largely on the structural parameter values. In contrast, the prediction ranges, when fixing structural parameters, become much wider by allowing the possibility that players may not observe or be partially informed about their own payoff shocks (columns under MCE Fully robust).

\section{Conclusion} \label{sec:conclusion}

In this paper, we have studied the identification, estimation, and counterfactual analyses of a common class of dynamic games for empirical analysis, assuming the researcher knows some minimal information available to players. To facilitate informationally robust econometric analysis of dynamic games, we have defined \emph{Markov correlated equilibrium}, a dynamic analog of Bayes correlated equilibrium, and studied its properties. We have discussed new challenges that arise in dynamic environments and proposed practical solutions. We have applied our proposed method to study the dynamic entry and exit decisions of the two largest coffee chains in the US.

We conclude this paper with a discussion of future research directions. First, further refinement of Markov correlated equilibrium (and thus Bayes correlated equilibrium) will attract greater interest from empiricists. The type of informational robustness offered by Markov/Bayes correlated equilibrium admits too many possibilities. Markov correlated equilibrium identified sets and counterfactual outcomes can span a wide range, which may not be empirically informative; \citet{Magnolfi2023} find similar results in their empirical application. Numerical examples from \citet{koh2023stable} and \citet{Magnolfi2023} show that even complete information correlated equilibrium identified sets can be large, indicating that correlation in actions can also weaken the identifying power substantially. In some settings, we may be able to reasonably exclude some information structures from consideration. Additionally, we may want to restrict attention to some equilibrium selection mechanisms to narrow down the possibilities. It will be fruitful to find a tractable approach that can accommodate further assumptions on the underlying information structure and equilibrium selection mechanism.

Second, estimation with Markov correlated equilibrium remains computationally challenging. A similar problem exists for Bayes correlated equilibrium since the analyst needs to run a grid search. Studying how to conduct computationally tractable estimation and inference in this class of partially identified models will be crucial for broadening the scope of applications.

\bigskip
\paragraph{Disclosure Statement} During the preparation of this work the authors used Open AI's ChatGPT 5.0 in order to check grammar and improve clarity of the writing. After using this tool/service, the authors reviewed and edited the content as needed and takes full responsibility for the content of the published article. Wharton Research Data Services (WRDS) was used in preparing part of the data set used in the research reported in this manuscript. This service and the data available thereon constitute valuable intellectual property and trade secrets of WRDS and/or its third-party suppliers. The authors declare that no financial or personal conflicts of interest influenced the research presented in this paper.

\appendix

\setcounter{table}{0}
\setcounter{figure}{0}
\renewcommand{\thetable}{\Alph{section}\arabic{table}}
\renewcommand{\thefigure}{\Alph{section}\arabic{figure}}

\section{Proofs}\label{sec:Proofs}

We use BNE, BCE, MPE, and MCE for Bayes Nash equilibrium, Bayes correlated equilibrium, Markov perfect equilibrium, and Markov correlated equilibrium, respectively, for brevity in the following proofs.

\subsection{Proof of Lemma \ref{lemma:one.shot.deviation.principle.formulation.of.Markov.perfect.equilibrium}} \label{sec:proof.of.lemma.1}
\begin{proof}
$(\Rightarrow)$ Let $\beta$ be a MPE of $(G,S)$. $\beta$ induces $(G^\beta,S)$. The BNE best-response condition is directly implied by the MPE condition of $(G,S)$. 

$(\Leftarrow)$ Let $\beta$ be a BNE of $(G^\beta,S)$. The BNE best-response condition implies that there is no profitable one-shot deviation given that each player expects all players to follow the prescription of $\beta$ in the future. The absence of profitable one-shot deviation implies that $\beta$ is a subgame perfect equilibrium of $(G,S)$ by the one-shot deviation principle. Therefore, $\beta$ is a MPE of $(G,S)$. 
\end{proof}
\subsection{Proof of Lemma \ref{Lemma:MCEandBCE}} \label{sec:proof.of.lemma.2}
\begin{proof}
The statement follows by applying the one-shot deviation principle.
\end{proof}
\subsection{Proof of Lemma \ref{lemma:identical.reduced.form.games}} \label{sec:proof.of.lemma.3}
\begin{proof}
It is enough to show that $V_{i}^{\beta}\left(x\right)=V_{i}^{\sigma}\left(x\right)$ for all $i\in\mathcal{I}$ and $x\in\mathcal{X}$ because then it follows that $v_{i}^{\beta}\left(a,x,\varepsilon\right)=v_{i}^{\sigma}\left(a,x,\varepsilon\right)$ for all $i\in\mathcal{I}$, $a\in\mathcal{A}$, $x\in\mathcal{X}$, and $\varepsilon\in\mathcal{E}$. 

Recall that $V_{i}^{\beta}$ and $V_{i}^{\sigma}$ satisfy \eqref{eqn:mpe.ex.ante.value.function} for $\left(G,S^{*}\right) $ and \eqref{eqn:ex.ante.value.function} for $\left(G,S\right) $, respectively. Since $\beta$ induces $\sigma$, \eqref{eqn:mpe.ex.ante.value.function} for $\left(G,S^{*}\right) $ is equivalent to
\begin{align*}
V_{i}^{\beta}\left(x\right)	& =\sum_{\varepsilon,\tau,\tilde{\tau},a}\psi_{\varepsilon\vert x}\pi_{\tau\vert x,\varepsilon}\lambda_{\tilde{\tau}\vert x,\varepsilon,\tau}\beta_{a\vert x,\tau,\tilde{\tau}}\left(u_{i}\left(a,x,\varepsilon\right)+\delta\sum_{x'}V_{i}^{\beta}\left(x'\right)f_{x'\vert a,x,\varepsilon}\right) \\
	& =\sum_{\varepsilon,\tau,a} \psi_{\varepsilon\vert x} \pi_{\tau\vert x,\varepsilon} \sigma_{a\vert x,\varepsilon,\tau} \left(u_{i} \left(a,x,\varepsilon\right) +\delta\sum_{x'} V_{i}^{\beta} \left(x'\right) f_{x'\vert a,x,\varepsilon}\right),
\end{align*}
where $\beta_{a\vert x,\tau,\tilde{\tau}}\equiv\prod_{i=1}^{I}\beta_{i}\left(a_{i}\vert x,\tau_{i},\tilde{\tau}_{i}\right)$. Comparison of the above equation for $V_{i}^{\beta}$ and \eqref{eqn:ex.ante.value.function} for $\left(G,S\right) $ that defines $V_{i}^{\sigma}$ implies that $V_{i}^{\beta}=V_{i}^{\sigma}$. 
\end{proof}
\subsection{Proof of Theorem \ref{thm:informational_robustness}} \label{sec:proof.of.theorem.1}

\begin{proof}
To prove the statement of the theorem, we make use of Theorem 1 of \cite{Bergemann2016} (BM), which we restate below.

\begin{lemma}[BM Theorem 1]
Let $(G,S)$ be a static game of incomplete information. A decision rule $\sigma$ is a BCE of $\left(G,S\right)$ if and only if, for some expansion $S^{*}$ of $S$, there is a BNE of $\left(G,S^{*}\right)$ that induces $\sigma$. 
\end{lemma}

$\left(\subseteq\right)$ Suppose $\sigma$ is a MCE of $\left(G,S\right)$. We want to show that there exists an expansion $S^{*}$ and a strategy profile $\beta$ in $\left(G,S^{*}\right)$ such that $\beta$ is a MPE of $\left(G,S^{*}\right)$ and $\beta$ induces $\sigma$. Since $\sigma$ is a MCE of $\left(G,S\right)$, it is a BCE of $\left(G^{\sigma},S\right)$. By BM Theorem 1, there exists an expansion $S^{*}$ of $S$ and a strategy profile $\beta$ of $\left(G^{\sigma},S^{*}\right)$ such that $\beta$ is a BNE of $\left(G^{\sigma},S^{*}\right)$ and $\beta$ induces $\sigma$. But since $G^{\sigma}=G^{\beta}$, $\beta$ is also a BNE of $\left(G^{\beta},S^{*}\right)$, which in turn implies that $\beta$ is a MPE of $\left(G,S^{*}\right)$.

$\left(\supseteq\right)$ Suppose $\beta$ is a MPE of $\left(G,S^{*}\right)$. We want to show that if $\beta$ induces $\sigma$ in $\left(G,S\right)$, then $\sigma$ is a MCE of $\left(G,S\right)$. Since $\beta$ is a MPE of $\left(G,S^{*}\right)$, it is a BNE of $\left(G^{\beta},S^{*}\right)$. By BM Theorem 1, if $\sigma$ is induced by $\beta$, then $\sigma$ is a BCE of $\left(G^{\beta},S\right)$. But since $G^{\beta}=G^{\sigma}$, $\sigma$ is also a BCE of $\left(G^{\sigma},S\right)$, which in turn implies that $\sigma$ is a MCE of $\left(G,S\right)$. 
\end{proof}
\subsection{Proof of Corollary \ref{cor:informational.robustness.ccp}} \label{sec:proof.of.corollary.1}
\begin{proof}
$\left(\subseteq\right)$ Take $\phi\in\mathcal{P}_{a\vert x}^{MCE}\left(G,S\right)$. By definition, there exists $q\in\mathcal{P}_{a\vert x,\varepsilon,\tau}^{MCE}\left(G,S\right)$ such that $q$ induces $\phi$, i.e.,$\phi_{a\vert x}=\sum_{\varepsilon\in\mathcal{E},\tau\in\mathcal{T}}\psi_{\varepsilon\vert x}\pi_{\tau\vert x,\varepsilon}q_{a\vert x,\varepsilon,\tau}$, $\forall a\in\mathcal{A},x\in\mathcal{X}$. Then there exists $S^{*}$ such that $S^{*}\succsim_{E}S$ and $q\in\mathcal{P}_{a\vert x,\varepsilon,\tau}^{MPE}\left(G,S^{*}\right)$, implying that $\phi\in\mathcal{P}_{a\vert x}^{MPE}\left(G,S^{*}\right)$.

$\left(\supseteq\right)$ Take $\phi\in\bigcup_{S^{*}\succsim_{E}S}\mathcal{P}_{a\vert x}^{MPE}\left(G,S^{*}\right)$. Then there exists $S^{*}$ such that $S^{*}\succsim_{E}S$ and $\phi\in\mathcal{P}_{a\vert x}^{MPE}\left(G,S^{*}\right)$, which implies that there exists $q\in\mathcal{P}_{a\vert x,\varepsilon,\tau}^{MPE}\left(G,S^{*}\right)$ such that $q$ induces $\phi$. Then since $q\in\mathcal{P}_{a\vert x,\varepsilon,\tau}^{MCE}\left(G,S\right)$, we have $q\in\mathcal{P}_{a\vert x}^{MCE}\left(G,S\right)$.
\end{proof}
\subsection{Proof of Theorem \ref{thm:more.info.smaller.mce}} \label{sec:proof.of.theorem.2}
\begin{proof}
    Let $\phi \in \mathcal{P}_{a \vert x, \varepsilon, \tau}^{MCE}(G,S^1)$. We want to show $\phi \in \mathcal{P}_{a \vert x, \varepsilon, \tau}^{MCE}(G,S^2)$. From our Theorem \ref{thm:informational_robustness}, we have
    \begin{align*}
        \mathcal{P}_{a \vert x, \varepsilon, \tau}^{MCE}(G,S^1) = \bigcup_{S^* \succsim_E S^1} \mathcal{P}_{a \vert x, \varepsilon, \tau}^{MPE} (G,S^*) \text{ and }
        \mathcal{P}_{a \vert x, \varepsilon, \tau}^{MCE}(G,S^2) = \bigcup_{S^{**} \succsim_E S^2} \mathcal{P}_{a \vert x, \varepsilon, \tau}^{MPE} (G,S^{**}).
    \end{align*}
    If $\phi \in \mathcal{P}_{a \vert x, \varepsilon, \tau}^{MCE}(G,S^1)$, then there exists $S^*$ such that $\phi \in \mathcal{P}_{a \vert x, \varepsilon, \tau}^{MPE} (G,S^*)$. But since $S^* \succsim_E S^1$ and $S^1 \succsim_E S^2$, we have $S^* \succsim_E S^2$, which implies $\phi \in \bigcup_{S^{**} \succsim_E S^2} \mathcal{P}_{a \vert x, \varepsilon, \tau}^{MPE} (G,S^{**})$, which is what we wanted to show.
\end{proof}
\subsection{Proof of Theorem \ref{thm:equivalence of identified sets}} \label{sec:proof.of.theorem.3}
\begin{proof}
We want to show that $\Theta_I^\mathit{MCE}(S) = \bigcup_{\tilde{S} \succsim_E S} \Theta_I^\mathit{MPE}(\tilde{S})$. Take $\theta \in \Theta_I^\mathit{MCE}(S)$. By definition, $\phi \in \mathcal{P}^\mathit{MCE}_{a \vert x} (G^\theta, S)$. By Corollary \ref{cor:informational.robustness.ccp}, there exists some $S^* \succsim_E S$ such that $ \phi \in \mathcal{P}^\mathit{MPE}_{a \vert x} (G, S^*)$. But this immediately implies $\theta \in \Theta_I^\mathit{MPE}(S^*)$, and thus $\theta \in \bigcup_{\tilde{S} \succsim_E S} \Theta_I^\mathit{MPE}(\tilde{S})$. The proof of converse follows the above steps in reverse.
\end{proof}
\subsection{Proof of Theorem \ref{thm:tighter_set}} \label{sec:proof.of.theorem.4}
\begin{proof}
The statement follows from Theorem \ref{thm:more.info.smaller.mce}.
\end{proof}
\subsection{Proof of Theorem \ref{thm:MPEC}} \label{sec:proof.of.theorem.5}
\begin{proof}
    The statement directly follows from the fact that the constraints \eqref{eqn:decision.rule}, \eqref{eqn:consistency.1}, \eqref{eqn:obedience.1}, and \eqref{eqn:value.2} characterize the MCE identification conditions as implied by the definition of identified sets. Note that the ex-ante value function equation \eqref{eqn:value.2} is equivalent to \eqref{eqn:ex.ante.value.function} because
    \begin{align*}
        \sum_{\varepsilon, \tau, a} \psi_{\varepsilon \vert x} \pi_{\tau \vert x, \varepsilon} \sigma_{a \vert x, \varepsilon, \tau} \sum_{x'} V_{i,x'} f_{x' \vert a, x} &= \sum_{a, x'} \left(\sum_{\varepsilon, \tau} \psi_{\varepsilon \vert x} \pi_{\tau \vert x, \varepsilon} \sigma_{a \vert x, \varepsilon, \tau} \right)  V_{i,x'} f_{x' \vert a, x} \\
        &= \sum_{a, x'} \phi_{a \vert x}  V_{i,x'} f_{x' \vert a, x}.
    \end{align*}
\end{proof}
\subsection{Proof of Theorem \ref{thm:fully.robust.identified.set}} \label{sec:proof.of.theorem.6}
\begin{proof}
    The ex-ante value function \eqref{eqn:fully.robust.value} can be derived by expanding \eqref{eqn:ex.ante.value.function} and imposing \eqref{eqn:fully.robust.consistency} similarly to the proof of Theorem \ref{thm:MPEC} in Appendix \ref{sec:proof.of.theorem.5}. The obedience condition \eqref{eqn:fully.robust.obedience.2} is derived as follows. The obedience condition in the fully robust case is
    \begin{equation}\label{eqn:fully.robust.obedience.11}
    \sum_{\varepsilon, a_{-i}} \psi_{\varepsilon \vert x} \sigma_{a \vert x, \varepsilon} \partial v_i^\theta( a_i', a_i, x, \varepsilon_i) \leq 0.    
    \end{equation}
    But since 
\begin{align*}
    & \sum_{\varepsilon,a_{-i}} \psi_{\varepsilon\vert x} \sigma_{a\vert x,\varepsilon} v_i^\theta (a_i',a_{-i},x,\varepsilon_i) \\
    = & \sum_{\varepsilon, a_{-i}} \psi_{\varepsilon \vert x} \sigma_{a \vert x, \varepsilon} \left(u_i^\theta (a_i',a_{-i},x,\varepsilon_i) + \delta \sum_{x'}V_{i,x'}f_{x'\vert a_i',a_{-i},x} \right) \\
 = & \sum_{\varepsilon,a_{-i}} \psi_{\varepsilon \vert x} \sigma_{a \vert x, \varepsilon} u_i^\theta(a_i',a_{-i},x,\varepsilon_i) + \sum_{a_{-i}} \left(\sum_{\varepsilon} \psi_{\varepsilon \vert x} \sigma_{a \vert x, \varepsilon} \right) \left(\delta \sum_{x'} V_{i,x'}f_{x' \vert a_i', a_{-i}, x} \right) \\
 = & \sum_{\varepsilon,a_{-i}} \psi_{\varepsilon \vert x} \sigma_{a \vert x, \varepsilon} u_i^\theta(a_i',a_{-i},x,\varepsilon_i) + \delta  \sum_{a_{-i},x'} \phi_{a\vert x} V_{i,x'} f_{x' \vert a_i', a_{-i}, x},
\end{align*}
where the last equation is obtained by imposing \eqref{eqn:fully.robust.consistency}, 
the obedience condition \eqref{eqn:fully.robust.obedience.11} can be replaced by 
\begin{equation*} 
\sum_{\varepsilon,a_{-i}} \psi_{\varepsilon \vert x} \sigma_{a \vert x, \varepsilon} \partial u_i^\theta(a_i',a,x,\varepsilon_i) + \delta \sum_{a_{-i},x'} \phi_{a\vert x} V_{i,x'} \partial f_{x' \vert a_i',a,x} \leq 0, \quad \forall i, x, a_i, a_i'
\end{equation*}
where $\partial u_i^\theta(a_i',a,x,\varepsilon_i) \equiv u_i^\theta(a_i',a_{-i},x,\varepsilon_i) - u_i^\theta(a,x,\varepsilon_i)$ and $\partial f_{x' \vert a_i', a, x} \equiv f_{x' \vert a_i', a_{-i}, x} - f_{x' \vert a, x}$. 
\end{proof}
\subsection{Proof of Theorem \ref{thm:confidence.set}} \label{sec:proof.of.theorem.7}

\begin{proof}
By construction, $\mathrm{Pr}\left(\Theta_I(\phi) \subseteq \bigcup_{\tilde{\phi} \in \Phi_\alpha} \Theta_I(\tilde{\phi}) \right) \geq \mathrm{Pr} \left(\phi \in \Phi_\alpha \right)$. By taking the limit inferior on both sides, we obtain the desired result. The remaining statement holds by the definitions of $\Theta_I(\phi) $ and $\widehat{\Theta}_I^\alpha$.
\end{proof}

\section{Results Without Exclusion Restriction \label{sec:results.wo.exclusion}}

\begin{table}[htb!]

\caption{\label{tab:estimation.results.woexclusion}Structural parameter estimates without exclusion restriction}
\centering
\resizebox{\linewidth}{!}{\begin{threeparttable}
\begin{tabular}[t]{lccc}
\toprule
 & \makecell{MPE\\Standard incomplete\\(1)} & \makecell{MCE\\Standard incomplete\\(2)} & \makecell{MCE\\Fully robust\\(3)}\\
\midrule
Intercept & $[0.120, 0.143]$ & $[0.114, 0.356]$ & $[-0.530, 0.390]$\\
 & $0.132$ & - & -\\
Eating places (log) & $[0.106, 0.114]$ & $[0.000, 0.164]$ & $[0.000, 0.368]$\\
 & $0.110$ & - & -\\
Entry cost & $[-8.871, -8.756]$ & $[-9.051, -5.792]$ & $[-9.087, -0.006]$\\
 & $-8.813$ & - & -\\
Starbucks competitive effect & $[0.077, 0.096]$ & $[-0.176, 0.146]$ & $[-0.677, 0.701]$\\
 & $0.086$ & - & -\\
Dunkin' competitive effect & $[-0.015, 0.005]$ & $[-0.195, 0.211]$ & $[-0.708, 0.632]$\\
 & $-0.005$ & - & -\\
\bottomrule
\end{tabular}
\begin{tablenotes}[para]
\footnotesize
\item \textit{Note:} Column (1) reports the 95\% confidence intervals (CIs) and point estimates for the model with the standard incomplete information Markov perfect equilibrium (MPE) assumption. Columns (2) and (3) use our Markov correlated equilibrium (MCE) framework with the standard incomplete information and without minimal signals, respectively. The parameter spaces of the eating places (log) and entry cost are bounded below and above by zero, respectively. CIs are bootstrapped with resampling at the market level.
\end{tablenotes}
\end{threeparttable}}
\end{table}

\begin{figure}[htb!]
    \centering
    \includegraphics[width=0.49\textwidth]{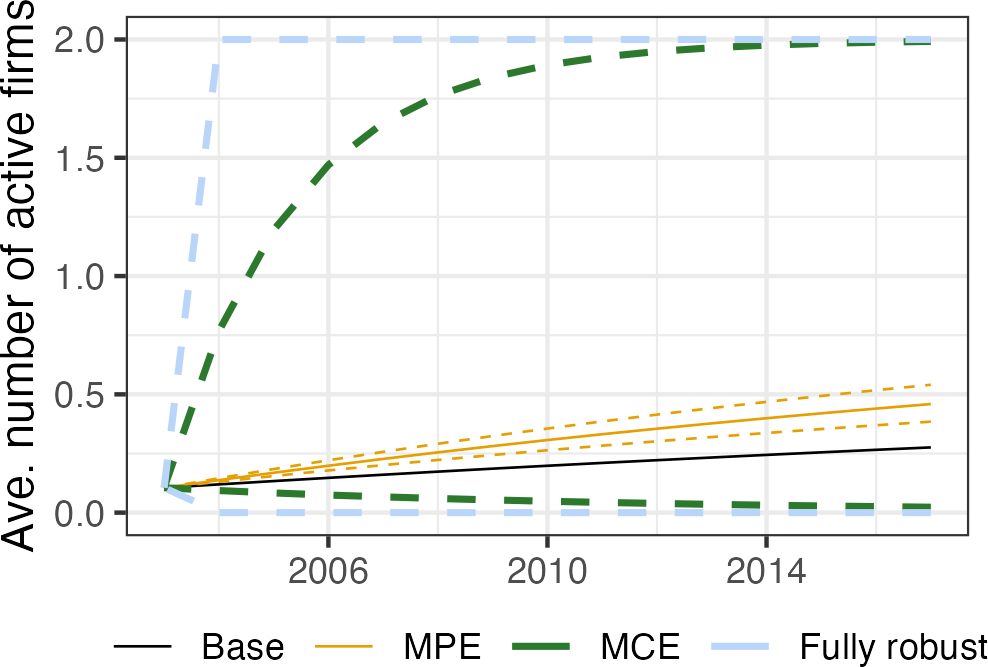}
    \caption{Counterfactual market status without exclusion restriction}
    \label{fig:counterfactual.wo.exclusion}
\end{figure}


\singlespacing
\bibliographystyle{econ}
\bibliography{ref}

\clearpage
\part*{Online Appendix}
\setstretch{1.3} 

\appendix

\setcounter{equation}{0}
\setcounter{table}{0}
\setcounter{figure}{0}
\renewcommand{\theequation}{\Alph{section}\arabic{equation}}
\renewcommand{\thetable}{\Alph{section}\arabic{table}}
\renewcommand{\thefigure}{\Alph{section}\arabic{figure}}

\section{Empirical Application Details} \label{sec:empirical.application.details}

\subsection{Interpretation of Counterfactual Analysis \label{sec:interpretation.of.counterfactual.analysis}}





As noted in a series of papers \citep*{aguirregabiria2014identification, kalouptsidi2017non, kalouptsidi2021identification}, normalizing the flow payoff from the outside option to zero is not without loss of generality for counterfactual analysis. However, we provide a way to interpret our normalization and counterfactual analysis under the assumption that a parametric specification is correct. 

Suppose that the flow payoff is given by
\begin{equation}\label{eqn:full.flow.payoff}
    u_i^\theta(a_{it}, a_{jt}, x_t, \varepsilon_{it}) = 
    \begin{cases}
    \theta_{i,0}^* + \theta_{i,w}^*w_{t} + \theta_{i,ec}^*(1-z_{it}) + \theta_{i,ce}^* a_{jt} + \varepsilon_{it} & \text{if $a_{it} = 1$} \\
    \theta_{i,sv}^* z_{it} & \text{if $a_{it} = 0$}
    \end{cases}
    ,
\end{equation}
where $\theta_{i,ec}^* < 0$ is the true entry cost and $\theta_{i,sv}^* >0$ is the true scrap value. A critical assumption in \eqref{eqn:full.flow.payoff} is that the scrap value is independent of the market characteristics. Payoff \eqref{eqn:full.flow.payoff} implies that the payoff difference is
\begin{align*}
&u_i^\theta(1,a_{jt},x_t,\varepsilon_{it}) - u_i^\theta(0,a_{jt},x_t,\varepsilon_{it}) =\\
& \hspace{30pt} \left(\theta_{i,0}^* - \theta_{i,sv}^* \right) + \theta_{i,w}^*w_t + \left( \theta_{i,ec}^* + \theta_{i,sv}^* \right) (1-z_{it}) + \theta_{i,ce}^* a_{jt} + \varepsilon_{it}.
\end{align*}
Thus, the estimated intercept captures $(\theta_{i,0}^* - \theta_{i,sv}^*)$ (i.e., fixed cost net of scrap value), and the estimated entry cost captures $(\theta_{i,ec}^* + \theta_{i,sv}^*)$ (i.e., entry cost net of scrap value). This is consistent with \citet{aguirregabiria2014identification} that the fixed cost, entry cost, and scrap value are not separately identified.

Now consider the counterfactual experiment of reducing the entry cost as in Section \ref{sec:counterfactual}. One can show that this is equivalent to holding $\theta_{i,sv}^*$ fixed and reducing $\theta_{i,ec}^*$ (making it less negative) to $\theta_{i,ec}^{**}$ by
\[
\theta_{i,ec}^{**} = \theta_{i,ec}^* - \lambda \underbrace{(\theta_{i,ec}^* + \theta_{i,sv}^*)}_{<0},
\]
where $\lambda$ is the proportion of entry cost intended to be cut. Thus, our counterfactual experiment is equivalent to providing an entry subsidy equal to $\lambda$ of the entry cost net of scrap value.

\subsection{Computation \label{sec:computation}}

\subsubsection{Estimation \label{sec:computation.estimation}}

We describe our estimation strategy using the excluded variable with large support. We estimate a single-agent discrete choice model with flow profit,
\begin{equation*}
    u_i^\theta(a_{imt},x_{mt},\varepsilon_{imt}) = a_{imt} \{\theta_{w}^\top w_{m} + \theta_{i,v} v_{im} + \theta_{ec}(1 - z_{imt}) + \varepsilon_{imt}\},
\end{equation*}
using markets where firm $i$ is the only potential entrant. Firm $j$ is not a potential entrant in a market if it had no outlets in the corresponding county during the sample period. The competitive effects term, $\theta_{i,ce} a_{jmt}$, is dropped from the two-player flow profit \eqref{eqn:empirical.model} since firm $j$ is not a potential entrant, i.e., $a_{jmt}$ is always $0$. We apply the two-step pseudo-likelihood method of \citet{Aguirregabiria2002} in the estimation. Table \ref{tab:single.logit.results} in Online Appendix \ref{sec:additional.estimation.results} shows the first-step logit model estimates. Table \ref{tab:estimation.results} columns (2) and (3) tabulate the structural estimates $(\hat{\theta}_{w},\hat{\theta}_{S,v},\hat{\theta}_{D,v},\hat{\theta}_{ec}) $, where $i = S $ for Starbucks and $i = D $ for Dunkin', in the rows from ``Intercept'' to ``Entry cost.''

We estimate the Markov correlated equilibrium identified set of the competitive effects parameters $\theta = (\theta_{S,ce},\theta_{D,ce}) $ in the flow profit \eqref{eqn:empirical.model} using markets where both firms are potential entrants, considering that the non-competitive effects parameters are given as those estimated from the single-player markets. We adopt the same non-competitive effects parameter estimates $(\hat{\theta}_{w},\hat{\theta}_{S,v},\hat{\theta}_{D,v},\hat{\theta}_{ec}) $ in the two Markov correlated equilibrium identified set estimations to demonstrate the identifying power of our Markov correlated equilibrium framework. We apply Theorem \ref{thm:confidence.set} to obtain the confidence region of the identified set of $\theta \in \Theta$, denoted as $\widehat{\Theta}_I$. We take the criterion function approach as commonly done in the partial identification literature (e.g., \citet*{Chernozhukov2007}) rather than testing whether $\theta \in \widehat{\Theta}_I $ by solving a feasibility program that directly follows Theorem \ref{thm:confidence.set}. We define a criterion function $Q: \Theta \mapsto \mathbb{R}_+$ such that $Q(\theta)=0$ if $\theta \in \widehat{\Theta}_I$ and $Q(\theta)>0$ otherwise. The criterion function is informative about the degree of violation in the constraints, while the output of a feasibility program is binary (``accepted'' or ``rejected''). Moreover, $\widehat{\Theta}_I$ can be numerically approximated by considering the sublevel set $\widehat{\Theta}_I(c) \equiv \left\{ \theta: \; Q(\theta) \leq c \right\}$ with a small threshold $c > 0$. 

We define the criterion function so that it measures the degree of violation in the obedience condition \eqref{eqn:obedience.1} for a given $\theta$:
\begin{align*}
Q(\theta) = \min_{q \geq 0,\sigma,V,\phi} \quad q
\end{align*}
subject to
\begin{align*}
    &\sigma_{a \vert x, \varepsilon, \tau} \geq 0 \text{ for each $a$ and } \sum_a \sigma_{a\vert x, \varepsilon, \tau} = 1, &&\quad \forall x,\varepsilon,\tau \\
    & \phi_{a\vert x} = \sum_{\varepsilon, \tau} \psi_{\varepsilon \vert x }\pi_{\tau \vert x, \varepsilon}  \sigma_{a \vert x, \varepsilon, \tau} \text{ and } \hat{L}_{a\vert x} \leq \phi_{a\vert x} \leq \hat{U}_{a\vert x} &&\quad \forall a,x \\
    &\sum_{\varepsilon, \tau_{-i}, a_{-i}} \psi_{\varepsilon | x} \pi_{\tau \vert x, \varepsilon} \sigma_{a \vert x, \varepsilon, \tau} \partial v_i^\theta(a_i',a,x,\varepsilon_i) \leq q, &&\quad \forall i, x, \tau_i, a_i, a_i' \\
    & V_{i,x} = \sum_{\varepsilon,\tau,a} \psi_{\varepsilon\vert x} \pi_{\tau\vert x, \varepsilon} \sigma_{a\vert x,\varepsilon,\tau } u_i^\theta (a,x,\varepsilon_i) + \delta \sum_{a,x'} \phi_{a \vert x} V_{i,x'} f_{x'\vert a,x}, &&\quad \forall i,x
\end{align*}
where $q \geq 0$ measures the degree of violation in the obedience condition, $\hat{L} $ and $\hat{U} $ are the lower and upper simultaneous confidence intervals of the conditional choice probabilities (CCPs) $\phi $, and the other notations are the same as they appear in Theorem \ref{thm:MPEC}.

In our empirical application, $\psi_{\varepsilon\vert x} = \psi_{\varepsilon} $ and $\pi_{\tau\vert x, \varepsilon} = \pi_{\tau\vert \varepsilon} $ since idiosyncratic shock $\varepsilon$ and signal $\tau$ are independent of state $x$. $\psi_{\varepsilon}$ represents the joint distribution of two independent shocks $(\varepsilon_S,\varepsilon_D)$ each drawn from the standard logistic distribution. We discretize the standard logistic distribution using the best 10-point approximation in \citet{Kennan2006}. For the standard incomplete information, $\pi_{\tau\vert \varepsilon} = 1 $ if $\tau = \varepsilon $ and $0 $ otherwise. For the fully robust specification, $\mathcal{T} $ is a singleton, and $\tau $ always takes that value regardless of $\varepsilon $. We construct the simultaneous confidence intervals of the CCPs with the plug-in sup-t implementation of \citet{montiel2019simultaneous}. We bootstrap the multinomial logit model shown in Table \ref{tab:logit.results} in Online Appendix \ref{sec:additional.estimation.results} at the market level to estimate the covariance matrix of the CCPs in the plug-in sup-t implementation.

We grid search $400 $ $(= 20^2) $ Halton draws in the 2-dimensional parameter space of $\theta $ and numerically approximate $\widehat{\Theta}_I$ by collecting $\widehat{\Theta}_I(c)$. Theoretically, we expect the standard incomplete information Markov correlated equilibrium identified set to contain the standard incomplete information Markov perfect equilibrium identified point, and the fully robust identified set to contain the standard incomplete information Markov correlated equilibrium identified set. We determine the criterion function threshold $c$ to make our estimation results conform to this theoretical prediction. We set $c$ for the standard incomplete information Markov correlated equilibrium identified set so that its confidence intervals (CIs) contain the CIs of the Markov perfect equilibrium estimates, and then set $c$ for the fully robust identified set so that its CIs contain the standard incomplete information Markov correlated equilibrium identified set CIs.

We also estimate the model without the excluded variable, where we drop the excluded variable term from the flow profit \eqref{eqn:empirical.model} as follows:
\begin{equation*}
    u_i^\theta(a_{imt},x_{mt},\varepsilon_{imt}) = a_{imt} \{\theta_{w}^\top w_{m} + \theta_{ec}(1 - z_{imt}) + \theta_{i,ce} a_{jmt} + \varepsilon_{imt}\}.
\end{equation*}
We grid search 1 million $(\approx 16^5) $ Halton draws in the 5-dimensional parameter space of $\theta = (\theta_{w},\theta_{ec},\theta_{S,ce},\theta_{D,ce}) $ and collect $\widehat{\Theta}_I(c)$ similarly to the Markov correlated equilibrium identified set estimation described above. Table \ref{tab:logit.results.exnone} in Online Appendix \ref{sec:additional.estimation.results} shows the multinomial logit model used in the CCP covariance matrix estimation.\footnote{Alternatively, one may find a projection interval of $\Theta_I$ by minimizing $p^\top \theta$ subject to the constraints above, where $p \in \mathbb{R}^{\dim (\theta)}$ is a direction vector (see, e.g., \cite*{kaido2019confidence}). The projection approach alleviates the need for a grid search over the entire parameter space but can be numerically difficult.}

\subsubsection{Counterfactual Equilibrium \label{sec:computation.equilibrium}}

For the low scenario, we find an equilibrium that minimizes the expected number of active firms for given counterfactual parameters $\tilde{\theta} $ by solving the following problem:
\begin{align*}
\min_{\sigma,V} \quad \frac{1}{\vert \mathcal{X} \vert} \sum_x \{0 \cdot \phi_{0,0 \vert x} + 1 \cdot (\phi_{0,1 \vert x} + \phi_{1,0 \vert x}) + 2 \cdot \phi_{1,1 \vert x} \}
\end{align*}
subject to
\begin{align*}
    &\sigma_{a \vert x, \varepsilon, \tau} \geq 0 \text{ for each $a$ and } \sum_a \sigma_{a\vert x, \varepsilon, \tau} = 1, &&\quad \forall x,\varepsilon,\tau \\
    & \phi_{a\vert x} = \sum_{\varepsilon, \tau} \psi_{\varepsilon \vert x }\pi_{\tau \vert x, \varepsilon}  \sigma_{a \vert x, \varepsilon, \tau} &&\quad \forall a,x \\
    &\sum_{\varepsilon, \tau_{-i}, a_{-i}} \psi_{\varepsilon | x} \pi_{\tau \vert x, \varepsilon} \sigma_{a \vert x, \varepsilon, \tau} \partial v_i^{\tilde{\theta}}(a_i',a,x,\varepsilon_i) \leq 0, &&\quad \forall i, x, \tau_i, a_i, a_i' \\
    & V_{i,x} = \sum_{\varepsilon,\tau,a} \psi_{\varepsilon\vert x} \pi_{\tau\vert x, \varepsilon} \sigma_{a\vert x,\varepsilon,\tau } u_i^{\tilde{\theta}} (a,x,\varepsilon_i) + \delta \sum_{a,x'} \phi_{a \vert x} V_{i,x'} f_{x'\vert a,x}, &&\quad \forall i,x
\end{align*}
where $0 \cdot \phi_{0,0 \vert x} + 1 \cdot (\phi_{0,1 \vert x} + \phi_{1,0 \vert x}) + 2 \cdot \phi_{1,1 \vert x}$ is the expected number of active firms at state $x \in \mathcal{X}$ for equilibrium CCPs $\phi_{a_1,a_2 \vert x} $, $\vert \mathcal{X} \vert $ is the cardinality of $\mathcal{X}$, and the other notations are the same as they appear in Theorem \ref{thm:MPEC}. The objective function takes an unweighted average over $x$ to obtain the (unconditional) expected number of active firms. The constraints, which are the same as Theorem \ref{thm:MPEC}, ensure that the resulting decision rule $\sigma $ satisfies the Markov correlated equilibrium conditions. The specifications of $\psi_{\varepsilon \vert x } $ and $\pi_{\tau \vert x, \varepsilon} $ are the same as the estimation, as explained in Online Appendix \ref{sec:computation.estimation}. We find an equilibrium that maximizes this optimization problem for the high scenario.

Let superscripts $l $ and $u $ indicate the lower and upper bounds of the 95\% CIs of the structural estimates, respectively, e.g., $\hat{\theta}_{ec}^l $ is the lower bound of the 95\% CIs of the entry cost parameter estimate. When we only consider the uncertainty of the competitive effects parameters, we set the counterfactual parameters to $\tilde{\theta} = (\hat{\theta}_{w},\hat{\theta}_{S,v},\hat{\theta}_{D,v},0.9 \cdot \hat{\theta}_{ec},\hat{\theta}_{S,ce}^l,\hat{\theta}_{D,ce}^l) $ in the low scenario under the counterfactual experiment of reducing the entry cost by 10\%. $0.9 \cdot \hat{\theta}_{ec}$ indicates a 10\% reduction of the original entry cost estimate $\hat{\theta}_{ec} $. We use the upper bounds of the 95\% CIs of the competitive effects parameters, $(\hat{\theta}_{S,ce}^u,\hat{\theta}_{D,ce}^u) $, instead of the lower bounds $(\hat{\theta}_{S,ce}^l,\hat{\theta}_{D,ce}^l) $ in the high scenario. To additionally incorporate the uncertainty of the non-competitive effects parameters, we set $\tilde{\theta} = (\hat{\theta}_{w}^l,\hat{\theta}_{S,v}^l,\hat{\theta}_{D,v}^l,0.9 \cdot \hat{\theta}_{ec}^l,\hat{\theta}_{S,ce}^l,\hat{\theta}_{D,ce}^l) $ and $\tilde{\theta} = (\hat{\theta}_{w}^u,\hat{\theta}_{S,v}^u,\hat{\theta}_{D,v}^u,0.9 \cdot \hat{\theta}_{ec}^u,\hat{\theta}_{S,ce}^u,\hat{\theta}_{D,ce}^u) $ in the low and high scenarios, respectively.

\subsection{Additional Estimation Results \label{sec:additional.estimation.results}}

\begin{table}[htb!]

\caption{\label{tab:single.logit.results}Binary logit parameter estimates in single-player markets}
\centering
\begin{threeparttable}
\begin{tabular}[t]{lcc}
\toprule
 & Coef. & 95\% CI\\
\midrule
Intercept & $-7.071$ & $[-7.288, -6.826]$\\
Eating places (log) & $0.935$ & $[0.880, 1.005]$\\
Starbucks outlets in county (log) & $0.170$ & $[0.135, 0.208]$\\
Dunkin' outlets in county (log) & $0.280$ & $[0.214, 0.354]$\\
Incumbent & $8.157$ & $[7.997, 8.317]$\\
\addlinespace
Number of Observations & 124,215 & \\
Log-likelihood & $-8583.7$ & \\
\bottomrule
\end{tabular}
\begin{tablenotes}[para]
\footnotesize
\item \textit{Note:} The table reports the coefficient estimates and 95\% confidence intervals (CIs) of the binary logit model of a single-player market having the firm active or not. The reference choice is the firm not active. CIs are bootstrapped with resampling at the market level.
\end{tablenotes}
\end{threeparttable}
\end{table}

\begin{table}[htb!]

\caption{\label{tab:logit.results}Multinomial logit parameter estimates in two-player markets}
\centering
\resizebox{\linewidth}{!}{\begin{threeparttable}
\begin{tabular}[t]{lccc}
\toprule
 & Starbucks & Dunkin' & Both\\
\midrule
Intercept & $-7.023$ & $-7.117$ & $-12.045$\\
 & $[-7.193, -6.878]$ & $[-7.276, -6.934]$ & $[-12.502, -11.719]$\\
Eating places (log) & $0.977$ & $0.632$ & $1.422$\\
 & $[0.945, 1.015]$ & $[0.598, 0.667]$ & $[1.334, 1.516]$\\
Starbucks outlets in county (log) & $0.184$ & $-0.223$ & $-0.159$\\
 & $[0.164, 0.205]$ & $[-0.242, -0.205]$ & $[-0.191, -0.128]$\\
Dunkin' outlets in county (log) & $-0.161$ & $0.417$ & $0.203$\\
 & $[-0.177, -0.146]$ & $[0.400, 0.438]$ & $[0.175, 0.228]$\\
Starbucks is incumbent & $8.365$ & $0.408$ & $8.644$\\
 & $[8.283, 8.485]$ & $[0.114, 0.711]$ & $[8.440, 8.868]$\\
Dunkin' is incumbent & $0.795$ & $8.563$ & $8.369$\\
 & $[0.536, 1.043]$ & $[8.476, 8.644]$ & $[8.138, 8.603]$\\
\addlinespace
Number of Observations & 456,090 &  & \\
Log-likelihood & $-53066.5$ &  & \\
\bottomrule
\end{tabular}
\begin{tablenotes}[para]
\footnotesize
\item \textit{Note:} The table reports the coefficient estimates and 95\% confidence intervals (CIs) of the multinomial logit model of a two-player market having Starbucks active, Dunkin' active, both active, or both not active. The reference choice is both not active. CIs are bootstrapped with resampling at the market level.
\end{tablenotes}
\end{threeparttable}}
\end{table}

\begin{table}[htb!]

\caption{\label{tab:logit.results.exnone}Multinomial logit parameter estimates without exclusion restriction}
\centering
\resizebox{\linewidth}{!}{\resizebox{\linewidth}{!}{\resizebox{\linewidth}{!}{\resizebox{\linewidth}{!}{\resizebox{\linewidth}{!}{\begin{threeparttable}
\begin{tabular}[t]{lccc}
\toprule
 & Starbucks & Dunkin' & Both\\
\midrule
Intercept & $-7.095$ & $-6.606$ & $-12.407$\\
 & $[-7.186, -7.011]$ & $[-6.690, -6.513]$ & $[-12.797, -12.062]$\\
Eating places (log) & $1.067$ & $0.640$ & $1.453$\\
 & $[1.035, 1.103]$ & $[0.599, 0.677]$ & $[1.352, 1.574]$\\
Starbucks is incumbent & $8.665$ & $0.491$ & $8.797$\\
 & $[8.588, 8.760]$ & $[0.215, 0.796]$ & $[8.579, 9.018]$\\
Dunkin' is incumbent & $0.731$ & $9.215$ & $8.937$\\
 & $[0.479, 0.980]$ & $[9.132, 9.297]$ & $[8.715, 9.153]$\\
\addlinespace
Number of Observations & 697,530 &  & \\
Log-likelihood & $-66013.2$ &  & \\
\bottomrule
\end{tabular}
\begin{tablenotes}[para]
\footnotesize
\item \textit{Note:} The table reports the coefficient estimates and 95\% confidence intervals (CIs) of the multinomial logit model of a two-player market having Starbucks active, Dunkin' active, both active, or both not active. The reference choice is both not active. CIs are bootstrapped with resampling at the market level.
\end{tablenotes}
\end{threeparttable}}}}}}
\end{table}



\end{document}